\documentclass[a4paper,fleqn,usenatbib]{mnras}
\usepackage[T1]{fontenc}
\usepackage{ae,aecompl}

\usepackage{times}
\usepackage{graphicx}	
\usepackage{amsmath}	
\usepackage{amssymb}	

\DeclareMathOperator{\sech}{sech}

\title[X-shaped structures in galaxies]{Measuring the X-shaped structures in 
edge-on galaxies}

\author[S. S. Savchenko et al.]{
S. S. Savchenko,$^{1}$\thanks{E-mail: s.s.savchenko@spbu.ru}
N. Ya. Sotnikova,$^{1}$
A. V. Mosenkov,$^{1,2}$
V. P. Reshetnikov$^{1}$
\newauthor
and D. V. Bizyaev$^{3,4}$
\\
$^{1}$St. Petersburg State University, 7/9 Universitetskaya nab., St.Petersburg, 199034 Russia \\
$^{2}$Sterrenkundig Observatorium, Universiteit Gent, Krijgslaan 281 S9,
B-9000 Gent, Belgium \\
$^{3}$Apache Point Observatory and New Mexico State University, Sunspot, NM, 88349, USA\\
$^{4}$Sternberg Astronomical Institute, Moscow State University, Moscow, Russia
}

\date{Accepted 2017 July 13. Received 2017 July 6; in original form 2017 April 7}

\pubyear{2017}

\begin{document}
\label{firstpage}
\pagerange{\pageref{firstpage}--\pageref{lastpage}}
\maketitle

\begin{abstract}
We present a detailed photometric study of a sample of 22 edge-on galaxies with clearly 
visible X-shaped structures. We propose a novel method to derive geometrical parameters
of these features, along with the parameters of their host galaxies based on the multi-component photometric
decomposition of galactic images. To include the X-shaped structure into our
photometric model, we use the {\small IMFIT} package,
 in which we implement a new component describing the X-shaped structure. 
 This method is applied for a sample of galaxies with available SDSS and {\it Spitzer} IRAC 3.6 $\mu$m observations.
 In order to explain our 
 results, we perform realistic
$N$-body simulations of a Milky Way-type galaxy and compare the observed  and the
model X-shaped structures.
Our main conclusions are as follows: 
(1) galaxies with strong X-shaped structures reside in approximately the same local 
environments as field galaxies; 
(2) the characteristic size of the X-shaped structures is about 2/3 of the bar size;
(3) there is a correlation between the X-shaped structure size and its observed flatness:
the larger structures 
are more flattened; 
(4) our $N$-body simulations qualitatively confirm the observational results and
support the bar-driven scenario for the X-shaped structure formation.
\end{abstract}




\section{Introduction}
Galaxies with boxy/peanut-shaped (B/PS) bulges 
were found as a curious 
deviation from normal galaxies with 
spherical-shaped bulges by \citet{Burbidge1959}. It is known now
that  the B/PS features are rather common in galaxies: the modern estimations of their 
probability vary from 20\% \citep{Yoshino2015} up to 40-45\%
\citep{Lutticke,Laurikainen2016} among all disc galaxies
observed edge-on. \citet{Erwin2017} found that the frequency of these features in
disc galaxies strongly depends on the galaxy mass and Hubble type: 
the B/PS fraction is higher in massive S0-Sa galaxies than in the other types.

As \citet{Bureau2006} found for a sample of 30 edge-on galaxies observed in the $K$-band,
the galaxies with a B/PS bulge tend to have a more complex morphology than galaxies with
other bulge types. These galaxies harbour the centered or off-centered X-shaped
structures, secondary maxima of brightness along the major-axis, and spiral-like
structures. Their surface brightness profiles are also more complex than in the galaxies
with classical bulges: usually they have breaks in their profiles
\citep[Freeman Type II profiles,][]{Freeman1970}.

Infrared observations and photometry of the red clump stars in the Galactic bulge
revealed that our Galaxy also contains a central X-shaped structure  
\citep{Laurikainen2014,Nataf2015,Ness2016}. 
In the recent work by \citet{Ciambur2016b} the first attempt to
quantify the properties of B/PS-structures is made. They used the Fourier analysis
to describe the surface brightness distribution in the central regions of galaxies
and retrieved some parameters of the X-shaped structures, such as their sizes both in
radial and vertical directions with respect to the galactic midplane, the integrated
`strength', and the peak amplitude of the $B_6$ Fourier component.

A commonly accepted model of a B/PS feature formation is the buckling
instability of a galactic bar \citep{Pfenniger1991, Athanassoula2002}.
The initially flat bar buckles via the vertical resonance of stellar orbits and becomes 
thicker. The final observed X-shaped structure is  made with
a set of these resonance orbits \citep{Combes1990}. This scenario was 
first obtained in the N-body simulations by \citet{Combes1981}, who 
found that the B/PS-bulges were formed shortly after the bar formation. 
\citet{Patsis2002} studied the formation of the X-shaped structures
from the point of view of individual orbits and found a set of orbit
families, a combination of which appears to be X-shaped when viewed edge-on.
The evolution of the bar vertical shape including the development of the
X-shaped appearance as a result of a recurrent buckling, is demonstrated in \citet{Martinez-Valpuesta2006}.

Observational evidence of the relationship between X-shaped structures
and bars was provided by \citet{Bureau1999}, where the position-velocity diagrams
were used to show that almost all edge-on galaxies with the B/PS inner isophotes also host
a thick bar. These findings were confirmed by \citet{Chung2004}, who
considered 24 edge-on galaxies with X-shaped structures, and demonstrated
that 22 of them show clear kinematic bar signatures. Kinematic identification
of X-shaped structures for face-on galaxies was performed in \citet{Mendez-Abreu2008}.

Another argument for the bar driven scenario of X-shaped structures is cylindrical
rotation ($\partial v/\partial\left| z \right|\sim 0$) inside of these features \citep{Combes1990}.
While it is usually the case, \citet{Williams2011} have found a counterexample in
their study of five edge-on galaxies: the galaxy IC~4767 does not rotate cylindrically,
though it has peanut-shaped central isophotes. The dark halo with the reduced central
concentration is required to explain these observations \citep{Athanassoula2002}.

Another prominent morphological feature in the central regions of many 
disc galaxies is a barlense \citep{Laurikainen2011}.
The barlenses and X-shaped structures are considered to be physically the 
same phenomenon \citep{Laurikainen2017}. They have very similar 
colours, and both features are thick in the comparison with the thin bar 
\citep{Herrera2017}, which suggests that these structures are results of 
the bar secular evolution. The appearance of these structures depends 
on the galaxy orientation, central flux concentration, and the 
steepness of the rotation curve in the central regions
\citep{Laurikainen2017, Salo2017}.

Although historically these structures are called B/PS-bulges, 
the  formation mechanism described above suggests that they have little 
in common with bulges. In this regard, following 
\citet{Patsis2002}, we avoid using the word ``bulges'' and refer 
to them as the ``X-shaped structures'' (or just ``X-structures'') instead.

The vertical structure of these features is more profound if a galaxy 
is seen at the edge-on orientation. However, even in this case an 
automated retrieval of the parameters of X-shaped structures is a challenging 
task because the light from the bulge and the disc dominate in the 
central regions of the galaxy. One approach to solve this problem is 
to enhance the appearance of the X-shaped structure by some image processing
techniques 
\citep[e.g. using the unsharp masking algorithm presented by][]{Bureau2006}.

In this work we propose a new approach to the estimation of the X-structure parameters.
We describe an algorithm which  allows us to derive geometrical parameters of the
X-structure, along  with the parameters of their host galaxies, via a photometric 
decomposition procedure. 
This allows us to separate the light of the X-structure from the 
light of the host galaxy. Once the model image of the host galaxy is 
obtained, it can be then subtracted from the galaxy image, leaving the 
X-structure clearly visible.

To interpret the results of our method, 
we perform the $N$-body simulations of a disc 
galaxy, in which we clearly observe the emergence of an X-structure. Then we compare 
the observed X-structure parameters with those obtained in our 
$N$-body simulations. 

The structure of the paper is organized as follows. 
In Sect.~2, we describe our sample of galaxies with visible 
X-structures. 
In Sect.~3, we outline the algorithm which we apply for the selected 
sample. 
The results of our investigation are shown in Sect.~4.
In Sect.~5, we describe the method of constructing our $N$-body models
that produce the X-structures, and analyze the obtained results. 
General conclusions are given in Sect.~6.

\section{The sample}
Our sample of galaxies with the X-shaped structures is drawn from the EGIS catalogue \citep{Bizyaev2014}, which 
contains 5747 edge-on galaxies. To form the sample of galaxies with prominent 
X-structures, we
visually inspected SDSS\footnote{\url{http://www.sdss.org/dr12/}} images of the EGIS galaxies.
We considered all (2021) galaxies with diameter estimated within the limits at the level of signal-to-noise of 2
per pixel in the $r$ band larger than 1 arcmin, and selected galaxies with noticeable central X shapes.
Only the galaxies with X-shaped structures are included in our sample. Although galaxies with 
central boxy regions represent the same structures viewed end-on \citep{Combes1981}, we do not consider
these objects in our work.

The final sample of the galaxies with X-shaped structures comprises 
151 objects in total. From our visual inspection, only a small part of the galaxies in the sample 
show no dust lanes. The X-structure may not be clearly seen due to the dust attenuation, and/or the orientation of
the bar towards the observer, and/or a small angular size of the X-structure
in a majority of the inspected objects. 

The fact that only 7.5\% out of the 2021 EGIS galaxies
are selected as the galaxies with X-structures 
indicates that only the brightest, well-seen X-structures were identified.

While a prominent dust lane can signify whether a galaxy is viewed perfectly edge-on, it
also severely affects derived structural parameters of edge-on galaxies. The influence of
 dust is significantly diminished when going 
from the optical SDSS bands to the near-infrared (NIR) domain. 
To study parameters of the X-structures in a NIR band, we added 
images of three galaxies (IC~2531, NGC~4013, and NGC~5529) 
observed at the 3.6 $\mu$m with the {\it Spitzer} infrared telescope 
(IRAC, \citealp{Fazio2004}). 
IC~2531 does not fall into the SDSS DR12
footprint, whereas the other two galaxies do, but their discs and 
X-structures are severely distorted by the dust such that only small tips of
their X-structures are visible above their dust lanes.

When we applied our method for determining the X-structure parameters (see Sect.~\ref{sec:methods})
to the selected sample, we found out that for many galaxies the decomposition procedure
(Sect.~\ref{sec:decomposition}) does not converge to a robust model of the X-structure.
There are several reasons for this.
Perhaps, the most common issue is a small
apparent size of the galaxy, and especially of the X-structure. Even if the X-structure is
clearly detected by eye, it often covers too few pixels in the image to be fitted 
automatically.
In many galaxies the X-structures are strongly obscured by the
dust lane such that only their tips are visible. These X-structures are also easy to detect
by eye but very difficult to process automatically, especially when the dust lane
appears to be clumpy.

The final sample of galaxies for which our analysis was successful consists of 19 galaxies
with the optical SDSS photometry,
and of 3 more galaxies with available IRAC 3.6$\mu$m observations.
Thus, our sample cannot be considered as complete 
since it is likely biased towards bright massive galaxies, in which the most prominent X-structures are detected.
We will refer to this sample of 22 galaxies as to the subsample.

Table~\ref{tab:sampleParams} summarizes some general properties of the subsample galaxies.
The morphological type $T$ and apparent diameter of the 25-th isophote in the $B$-band are taken from the
HyperLEDA\footnote{http://leda.univ-lyon1.fr/} database \citep{Makarov2014}.
The redshift values are taken from
the
NED\footnote{NASA/IPAC Extragalactic Database, \url{https://ned.ipac.caltech.edu/}} database.
We computed absolute magnitudes in the $r$ band from our photometric models
(see Sect.~\ref{sec:decomposition}) using the luminosity distance and
Galactic extinction values from the NED. The colours were computed from
our models and corrected for the extinction in the Galaxy.
Utilizing the decomposition models instead of the original images
for the galaxy flux estimation allows one to exclude 
the contamination from background objects, and also solves the problem of the
aperture size for
performing the photometry: one can integrate an analytical model over a wide range
of spatial coordinates to retrieve a precise value of the flux. The range of the
integration was set to be 10 times  the major axis of the 25-th isophote, to make
sure that the total flux outside the integration boundaries is negligible.

The average values of the type and the absolute magnitude are $\langle T
\rangle = 2.3 \pm 1.8$, $\langle M \rangle = -20.9 \pm 0.6$, which suggests
that our subsample consists of bright early-type spirals. The average observed
colour of the galaxies  ($\langle g - r \rangle = +0.86 \pm 0.07$) is also
consistent with edge-on early-type spirals (e.g., \citealt{Bizyaev2014}).

\begin{table}
  \caption{General parameters of the subsample galaxies with central X-structures} 
  \label{tab:sampleParams}
  \begin{tabular}{lcrclc}
    \hline
    Name & $T $ & $M_\mathrm{r}$  & $g-r$ & $d_{25}$ & $z$\\
         &   & mag   &  mag  & arcmin\\
    \hline
    PGC~24926 & 2.0  & -19.36 & 0.89 & 1.70 & 0.0054 \\
    PGC~26482 & 3.1  & -20.66 & 0.90 & 1.17 & 0.0287 \\
    PGC~28788 & 0.5  & -20.35 & 0.74 & 0.85 & 0.0178 \\
    PGC~28900 & 0.9  & -21.55 & 0.81 & 1.00 & 0.0214 \\
    PGC~32668 & 0.0  & -20.59 & 0.80 & 1.04 & 0.0215 \\
    PGC~37949 & 1.4  & -20.82 & 0.93 & 1.32 & 0.0210 \\
    PGC~39251 & -0.9 & -20.73 & 0.73 & 1.69 & 0.0067 \\
    PGC~44422 & 0.4  & -21.13 & 0.73 & $0.72^*$& 0.0256 \\ 
    PGC~69401 & 1.0  & -21.97 & 0.84 & 1.10 & 0.0312 \\
    PGC~34913 & 3.0  & -20.89 & 0.90 & 1.91 & 0.0146 \\
    PGC~45214 & 1.0  & -21.41 & 0.95 & 0.89 & 0.0406 \\
    PGC~10019 & 3.0  & -20.96 & 0.82 & 0.78 & 0.0362 \\
    PGC~30221 & 3.0  & -20.44 & 0.93 & 1.12 & 0.0209 \\
    PGC~55959 & 2.0  & -21.38 & 0.91 & 1.02 & 0.0322 \\
    ASK~361026.0&4.6  & -21.77 & 0.89 & $0.44^*$ & 0.0463\\ 
    PGC~53812 & 3.9  & -21.51 & 0.92 & 1.10 & 0.0359 \\
    PGC~02865 & 5.8  & -20.01 & 0.86 & 1.78 & 0.0170 \\
    PGC~21357 & 3.3  & -21.06 & 0.91 & 1.15 & 0.0393 \\
    PGC~69739 & 5.2  & -20.92 & 0.94 & 2.00 & 0.0242 \\
    \hline
    IC~2531    & 5.0  & -22.49 &  --  & 6.60 & 0.0082 \\
    NGC~4013   & 3.1  & -20.75 & 1.54 & 4.90 & 0.0027 \\
    NGC~5529   & 5.1  & -22.23 & 0.85 & 5.75 & 0.0096 \\
    \hline
  \end{tabular}
  \setcounter{table}{1}\\
  The columns show the name, morphological type, absolute magnitude in the SDSS $r$ band,
  extinction-corrected colour (g-r), and the diameter $d_{25}$ from HyperLEDA. The last
  column shows the redshift from the NED database.\\
   $^*$: there is no diameter value in the HyperLEDA database, the values were taken from
   the NED database.
\end{table}

\section{Methods}
\label{sec:methods}
In this section we describe the methods we used to extract parameters 
of the observed X-structures and their host galaxies. 
Our pipeline consists 
of three main steps: (i) preparation of galaxy images and generation of object masks, (ii) photometric
decomposition of the final galaxy images, (iii) and subsequent measurement of the X-structure parameters. 

\subsection{Images preparation}
We retrieved images of the subsample galaxies in the $r$ band data 
from SDSS DR12
(\citealt{Eisenstein2011}).
Since some galaxies occupy more than one SDSS field (for example, if 
a galaxy is big enough or located close to the field edge) we used the 
{\small SWARP} package \citep{Bertin2002} to concatenate adjacent fields.

The second step was to determine and subtract the sky background from the images. The SDSS 
pipeline includes a background subtraction routine, however we decided 
to perform our own background estimation to check for possible 
inaccuracies of the initial background subtraction. The algorithm
of the background estimation was as follows. 
First, using the
{\small SEXTRACTOR} code \citep{Bertin1996}, we create a map of pixels which are not occupied by any object
in the image. Then we fit the background intensity by a two-dimensional polynomial of the
first order using only unmasked pixels. The weights of
individual pixels in the fit are set to be the distance from the pixel to
the nearest object. This allows us to minimize the influence of possible
unmasked faint wings of extended objects.

After that, we rotated the images to align the galactic major axis along 
the $x$-axis. To estimate the position angle of the galaxy, we used the
method described in \citet{Martin-Navarro2012}. The main idea of the
method is to measure the number of galaxy pixels which lie inside of a
strip of a given width as a function of the position angle.
The optimal value of the position angle is the one which gives the highest
number of such pixels.

The next step was creating the point spread function (PSF) for every 
image of our subsample. Atmosphere blurring and telescope optics 
(to a less degree) can severely affect measured structural 
properties of galaxies \citep{Trujillo2001}. To compensate
the effects of seeing during the decomposition process (see below), 
one has to take into account the PSF of the image. In this paper 
we adopt the Moffat \citep{Moffat1969} function as an analytical 
approximation to the observed PSF. To retrieve the parameters of the Moffat 
function, we selected several good (not crowded, bright, but not 
saturated) stars in the image and fitted them with the Moffat 
function. Then these fit parameters of individual stars were 
averaged to obtain robust values of the adopted PSF model.

The last step was creating the mask images to mask out background and foreground 
objects, the light of which can affect the observed properties 
of the galaxies. To create the mask images, we used catalogues of objects
created with the {\small SEXTRACTOR} package \citep{Bertin1996} with subsequent 
manual inspection of the created masks.

The IRAC 3.6$\mu$m images  of NGC~4013 and NGC~5529 were downloaded from the
Spitzer Survey of Stellar Structure in Galaxies
archive\footnote{http://irsa.ipac.caltech.edu/data/SPITZER/S4G/}
\citep[S4G,][]{Sheth2010}, along with the masks and weight (one-sigma) images. The sky
background has already been subtracted from these frames. For IC~2531 we used the
observations available through the Spitzer Heritage Archive (the mosaic {\it*maic.fits}
and uncertainties {\it*munc.fits} files from the post-Basic Calibrated Data). The sky
background subtraction, rotation, cut out from the initial frames,
and masking were carried out in a similar way as the SDSS frames. 
To obtain PSF kernels for these three
galaxies, we used in-flight point response function (PRF) images for the centre of the IRAC 3.6$\mu$m
field\footnote{http://irsa.ipac.caltech.edu/data/SPITZER/docs/irac/calibrationfiles/} downsampled to the
0.6 (IC~2531) or 0.75 (NGC~4013 and NGC~5529) arcsec/pixel scale and re-rotated to correspond to the
analysed galaxy frame (the typical PSF FWHM for the IRAC 3.6$\mu$m images is 1.66 arcsec).

\subsection{The Photometric Decomposition}
\label{sec:decomposition}
The decomposition of a galaxy image is a widely used approach to 
obtain structural and photometric parameters of a galaxy. 
To perform the decomposition one has at first to construct a proper analytical 
model of the analyzed galaxy image and then find optimal parameters 
of this model via some optimisation procedure. The fit model should reflect 
the observed structure of the galaxy. It usually consists of separate 
components, which represent structural components of the galaxy. 
Below we describe the model which we adopt in this work and the algorithm 
to find optimal parameters of the model.

\subsubsection{The model}
\label{sec:model}
The main components of our model are a bulge, a disc, and 
an X-structure. 
The surface brightness distribution $I(r)$ in the bulge is modeled with the S\'ersic function 
\citep{Sersic1968}:
\begin{equation}
  I(r) = I_\mathrm{e}\exp \left( -\nu_\mathrm{n} \left[ 
  \left(\frac{r}{r_\mathrm{e}}  \right)^{1/n}-1 \right] \right) \, ,
\label{eq:sersic}
\end{equation}
where $I_\mathrm{e}$ is the effective surface brightness, 
$r_\mathrm{e}$ --- the effective radius, $n$ --- the 
S\'ersic parameter, 
and $\nu_\mathrm{n}$ is a function depending on $n$ \citep{Caon1993}.

The surface brightness of the disc in the ``edge-on'' orientation 
is usually represented by a two-dimensional function with the vertical 
and radial scale lengths. In the literature, one can find somewhat 
different approaches to represent this function. In this work we adopt it 
in the form consistent with that presented by~\citet{Erwin2015}:
\begin{equation}
  I(r, z) = 
  2hL_0\frac{r}{h}K_1 \left( \frac{r}{h} \right) 
  \sech^{2/m} \left( \frac{mz}{2z_0} \right) \, ,
  \label{eq:eondisc}
\end{equation}
where $h$ is the radial scale length, $z_0$ is the vertical scale 
height, $K_1$ is the modified Bessel function, and $L_0$ is the central 
luminosity density. The observed central surface brightness is 
related to $L_0$ as $I_0=2hL_0$. The last parameter is $m$, which is 
often set to a constant value of 1, but in this work we 
set it as a free parameter of the model. Note, that with 
$m\to \infty$ the vertical light distribution approaches 
the pure exponential shape (double exponential disc). Taking into account this fact, we set the 
upper limit on $m$ to be 20.0 since the further increasing of
this parameter does not affect the light distribution significantly:
the difference between the disc with $m=20$ and the disc with a pure
exponential vertical distribution is $\sim 10^{-3}$ in terms of the
total sum or residuals, therefore  we can consider the discs with $m=20$ as ones with the
exponential vertical profile.

To describe the observed surface brightness of an X-shaped structure, we 
use the modified Ferrers profile \citep{Ferrers1877}:

\begin{equation}
  I(r) = I_0 \left[ 1- \left( \frac{r}{r_{\mathrm{out}}} 
  \right)^{2-\beta} \right]^\gamma, \quad r\le r_{\mathrm{out}} \, .
  \label{eq:ferrer}
\end{equation}

The Ferrers profile has four free parameters: 
the outer radius $r_{\mathrm{out}}$ and 
the central intensity $I_0$, along with two numerical values 
$\beta$ and $\gamma$, which control the overall shape of the profile. 
The Ferrers profile has a central plateau with a steep decrease 
of the intensity in the outer regions, and it is often used to describe 
bars \citep[see e.g,][]{Laurikainen2007}.

To add an X-shaped appearance to the Ferrers profile, we 
modulate it with the $m=4$ Fourier mode:
\begin{equation}
  \label{eq:fourierm}
  r = r_0 \left( 1+ a_4 \cos \left( 4(\phi+\phi_0) \right) \right)\, .
\end{equation}
Here, the coefficient $a_4$ regulates the strength of the X-shaped distortion, and
$\phi_0$ defines its overall rotation.

These three components described above (the disc, the bulge, and 
the X-structure) are present in every photometric model of each subsample galaxy. 
Although we strived to restrict the number of the model components,
in some cases we had to introduce additional components (see below) when 
this simplified modelling was not satisfactory.

An important galaxy component is a dust structure, which is seen as a dimmed dust lane located close 
to the middle plane of the edge-on oriented galaxy. The simplest 
approach to account for the dust lane is to completely neglect it 
during the decomposition or to mask out the dust-reach region of the 
disc, however this can unpredictably affect the retrieved parameters of the decomposition.
A more sophisticated and precise way is to add an additional dust component and run a
radiative transfer code \citep[see e.g. {\tt FitSKIRT},][]{DeGeyter2014} to create a
self-consistent decomposition model, but this method is computationally consuming, especially
for a sample of several dozens of objects. In this paper, we take the dust lane into account
by adding it as a separate edge-on disc
\eqref{eq:eondisc} with \textit{negative} intensity
(in other words, the dust model is being subtracted from the overall 
model of the galaxy in contrast with the other ``regular'' components). 
We added this dust component only for those galaxies where a dust lane 
was clearly seen.

Some galaxies in our subsample (for example, PGC~28900 and PGC~39251) 
demonstrate two blobs within the disc at both sides of 
the X-structure. These blobs are not rare in galaxies
with X-shaped structures and arise in orbit simulations \citep{Patsis2002}. 
In our model, we simulated the presence of such blobs by adding 
an edge-on oriented, three-dimensional ring with a Gaussian distribution 
of the luminosity density along the radial and vertical directions. 
This model component has four free parameters: the ring
radius $a$, the luminosity density $J_0$ at this radius
and two Gaussian scales, radial and vertical. The observed two-dimensional 
brightness distribution is computed by the line-of-sight 
integration through the ring. When viewed edge-on, this model component
appears as two bright areas 
on both sides from the centre.

In order to simplify the decomposition of galaxies with additional components (a dust
lane or a ring), we split the whole procedure
into two steps. The first step is to find a relatively simple
"bulge+disc+X-structure" model. Then, in the second step, we add a
ring or a dust lane component utilizing parameters of the simple
model as initial conditions for the complex one.

We will refer hereafter to the galaxy model with all components 
except for the X-structure as a host model.

All components of the model, along with their main described above parameters, have a set of additional parameters which 
determines their appearance in the image. These parameters are coordinates
of the component centre $x_\mathrm{c}, y_\mathrm{c}$, ellipticity $e$, and position
angle $\mathrm{PA}$.

\subsubsection{The decomposition procedure}
The aim of the decomposition procedure is 
to find the best-fitting values of parameters of the multi-component model. These values 
can be found via optimisation algorithms, such as the
Levenberg-Marquardt gradient descent algorithm \citep{Levenberg1944} 
or the Nelder-Mead simplex algorithm \citep{Nelder1965}. The main 
issue here is to find proper initial parameters to pass them to
optimisation iterations. If initial values are not close enough to 
their optimal values, the optimisation process can converge to one of 
possible local minima or even to singular points of the model
function.

In this work we use the genetic algorithm (GA, \citealt{Goldberg1989}) 
to find the initial conditions for gradient descent method. GA is a 
computational algorithm for searching solutions of an optimisation 
problem which simulates the natural selection process.
The main idea of GA is to represent possible solutions of the 
problem as ``organisms'' with ``genes'' to be free parameters of a 
model. During the optimisation procedure these organisms evolve through 
the ``breeding'' and ``mutation'' processes to find a gene combination
that minimises the fitted function:
\begin{equation}
  \label{eq:fitness}
  f(g_1,g_2,...) = \sum_{i,j}w_{i,j} 
  \left[I_{\mathrm{obs}}(i,j) - I_{\mathrm{model}}(i,j;g_1,g_2,...)
  \right]^2 \, ,
\end{equation}
where $I_{\mathrm{obs}}(i,j; )$ is the observed intensity at the pixel 
with the coordinates $i$ and $j$, $I_{\mathrm{model}}(i,j;g_1,g_2,...)$ 
is the model intensity depending on the particular set of genes, 
and $w_{i,j}=\frac{1}{\sigma^2_{i,j}}$ are weights computed from 
per-pixel errors. 
The summation is performed over all unmasked image pixels. 
This function shows how close an organism to the solution
of the problem is, and is called the fitness function.

The fitness function (\ref{eq:fitness}) is computed for a number of
organisms (called a "generation"), and a subset of organisms is selected
based on the best values of the fitness function. These organisms are used
to create the next generation, and the cycling repeats until converging to optimal values.
The iterations start with a zero generation which consists of organisms with
randomly chosen parameter values.

In this work we set the generation size to be 250 organisms. 
The zero generation was twice as large. 
The iterations stop if the relative
decrease of the generation-wide average value of the fitness function is
less then $10^{-6}$ for five generations in a row.
To verify the quality of the obtained via GA solution, we use
it as an initial guess to the gradient descent algorithm. 
We adopt the results of the gradient descent algorithm as the 
final decomposition parameters. 

We use the flexible {\tt IMFIT} package  \citep{Erwin2015}
as a basis for the decomposition engine. 
We modified its source code to include the X-shaped 
Ferrers function, and wrote a {\tt PYTHON}\footnote{https://www.python.org} 
wrapper to parallelize computations for a large set of ``genetic" populations.

Estimating uncertainties of the parameters retrieved by this decomposition
procedure is a difficult task. The genetic algorithm does not provide such estimates,
and the gradient descent algorithm, which is used in the {\tt IMFIT} package, provides only the lower limits
of their values via the covariance matrix \citep{Erwin2015}. One way to find
the parameters uncertainties is to run the entire decomposition process several times.
Since the startup procedure of the genetic algorithm includes a random zero generation,
every run can converge to a slightly different model. One can use the scatter of the
parameters 
obtained in every rerun, 
to estimate the corresponding confidence intervals.

Unfortunately, this approach is computationally expensive, therefore we decided to apply it to just one galaxy of our sample,
in order to obtain typical values of uncertainties of the model parameters. We run the
decomposition procedure 50 times for PGC~24926 and retrieved the following relative 
standard deviations of the parameters: $\delta(\mu_\mathrm{e})=13\%$, $\delta(r_\mathrm{e})=17\%$,
$\delta(n)=11\%$, $\delta(\mu_0)=19\%$, $\delta(h)=2\%$, $\delta(m)=74\%$, $\delta(z_0)=4\%$
(only structural parameters given in Table~\ref{tab:decParams} are listed).
One can see that the highest uncertainty is for the parameter which governs the vertical
shape of the disc, $m$. This can be caused by the dust attenuation and/or insufficient resolution of the
vertical structure of this galaxy to reliably fit this parameter.

\subsection{The X-structure analysis}
The last step in the galaxy analysis is deriving parameters of its 
X-structure. To reduce the influence of the other galactic components 
(the disc, the bulge, etc. ) on the measured properties of the X-structure, we 
subtract their models obtained during the decomposition from the 
original galaxy image. The residual image represents the X-structure. 

\begin{figure}
  \center
  \includegraphics[width=0.75\columnwidth]{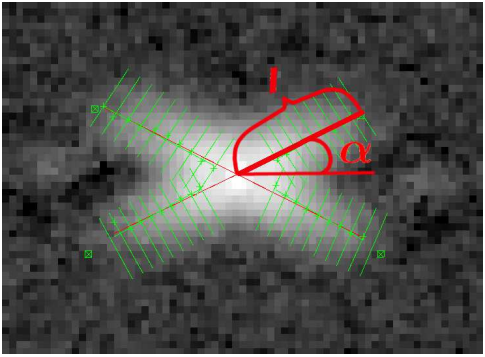}
    \caption{An example of the X-structure analysis. The thin lines perpendicular to the X-structure rays represent photometric cuts. 
      The ``+'' sign on each slice shows the maximum location along 
      the slice. 
      The long thick lines show the linear 
      fit to these points. 
      The x-in-a-box signs indicate 
      the limits of the fitting
       (see text).
      The estimated parameters $l$ and $\alpha$ are shown in the image
      for one ray.}
    \label{fig:xAnalisys}
\end{figure}

After that, for each ray of the X-structure we make a set of 
photometric slices perpendicular to the ray direction and fit them 
with a Gaussian function in order to find the location of the maximum 
intensity points on the ray. Then we  trace their location with
a straight line, to obtain the mean direction of 
the ray. In the last step, we derive the length of the ray ($l$) as 
a distance between the galaxy centre and the point along the ray where 
its intensity becomes fainter then 
$I_{\mathrm{b}} + 3\sigma_{\mathrm{b}}$, where 
$I_{\mathrm{b}}$ and $\sigma_{\mathrm{b}}$ are the background 
intensity and its 
uncertainty, respectively (this threshold roughly 
corresponds to the surface brightness level of 24
mag arcsec$^{-2}$
in the $r$ band). Fig.~\ref{fig:xAnalisys} demonstrates a result of
our analysis for PGC~32668. Another parameter, which we obtain from 
this analysis, is the angle $\alpha$ between the ray and the galactic 
midplane. 

Since the X-structure has four such rays, we performed this analysis 
for every ray separately and set the final parameters of the 
X-structure to be the mean values among all four rays.

Using the length of the 
obtained X-structure rays has a significant advantage in the comparison with utilizing
the parameters of the Ferrers function. The Ferrers function does not contain any
characteristic scale length, such as $r_\mathrm{e}$ for the S\'ersic profile and $h$ for
the exponential disc. The parameter $r_{\mathrm{out}}$ shows where the surface brightness
becomes exactly zero, but since the overall shape of the light curve is governed by the
two power-law values $\beta$ and $\gamma$, it is not evident where it
goes through a given intensity level.
Also, it is not  defined how to compare the sizes of the Ferrers components with
different $\beta$ and $\gamma$. On the other hand, the length of an X-structure ray
measured  in photometric profiles has a straightforward meaning: the length is measured 
down to a given intensity level along the ray. We should stress here that the Ferrers
component in our model, which describes the X-structure, is added
to perform more reliable overall photometric decomposition of the galaxy. Using this
approach, we are able to better fit the major structural components of the galaxy and,
at the same time, to better trace the X-structure, defined as the residual between the
galaxy image and the major model components.

\section{Results}
\subsection{The galactic environment}
Before going to the decomposition analysis, we need to investigate the 
spatial environment of the galaxies with the X-structures. For this purpose, 
for the 151 galaxies in our total sample
we estimate the number 
of neighbour galaxies using the SDSS DR12 database (\citealt{Eisenstein2011}). 

We define a galaxy
as a neighbour if it satisfies a set of requirements on its 
apparent angular distance from the target galaxy, the difference of 
their magnitudes, and radial velocities. To obtain more robust results, two 
different sets of such requirements were used: 
\begin{enumerate}
\item $\Delta r < 15 r_{\mathrm{petro}}$, $\Delta m < 2$mag, $\left|\Delta v\right| < 300$ km/s\,,
\item $\Delta r < 30 r_{\mathrm{petro}}$, $\Delta m < 2.5$ mag, $\left|\Delta v\right| < 300$ km/s\,,
\end{enumerate}

where $r_{\mathrm{petro}}$ is the $r$ band Petrosian radius of a target galaxy
and $\Delta v$ is the difference of radial velocities. The values of 
$r_{\mathrm{petro}}$ and $m$ are taken in the $r$ band. 
We will call these two sets of parameters for searching for neighbour galaxies as test 1 and test 2 sets.

Since the SDSS spectroscopic database is limited to $m_\mathrm{r}\approx 18\,\mathrm{mag}$ (\citealt{Strauss2002}),
we need to restrict our samples to the magnitude $m_\mathrm{r}=16\,\mathrm{mag}$ for the first set
of parameters, and $m_\mathrm{r}=15.5\,\mathrm{mag}$ for the second one, otherwise the limiting magnitude for the
neighbours search will exceed $18 \,\mathrm{mag}$ and the search will be performed in an incomplete 
domain of SDSS. After
this restriction, our sample of galaxies with X-structures was reduced to
117 objects for the test 1 and 92 objects for the test 2.

The results of the described spatial environment test can be 
affected by some parameters of the galaxies:
their apparent magnitudes and redshifts (due to incompleteness of the spectroscopic data for the faint end of the
sample), and angular sizes (since the search range depends on them). In order to take these issues into
account, the comparison sample should contain galaxies with similar distributions of these parameters.

The creation of the comparison sample was as follows. At the first step we selected 
all edge-on galaxies from the Galaxy Zoo \citep{Lintott2008} using the following
restrictions on votes: $P_{\mathrm{EDGE}} > 0.95$,
$P_{\mathrm{EL}} < 0.05$, $P_{\mathrm{CW}} < 0.05$ and $P_{\mathrm{ACW}} < 0.05$, which means 
that only objects with
high probability of being edge-on galaxy and low probability of being elliptical or non edge-on disc galaxy
were selected. Then, for every galaxy in our main sample 
with X-structures we found all edge-on galaxies
with similar size ($\Delta R{\mathrm{petro}}<20\%$), magnitude ($\Delta m_\mathrm{r} < 0.2 \,\mathrm{mag}$) and
redshift ($\Delta z < 0.01$). Next, we checked images of all these similar galaxies 
visually and removed those that demonstrate signs of being X-shaped or show boxy 
central isophotes.

After this step, for every galaxy with the X-structure from our sample, we have got a set of galaxies with similar
parameters, but without apparent X-shaped structures. To finally form the comparison sample, we randomly
picked up one galaxy from every set. As a result, we obtained the comparison sample with similar
distributions by angular size, apparent magnitude, and redshift and with the same size as our
sample of galaxies with the X-structures.

\begin{figure}
  \center
  \includegraphics[width=0.85\columnwidth]{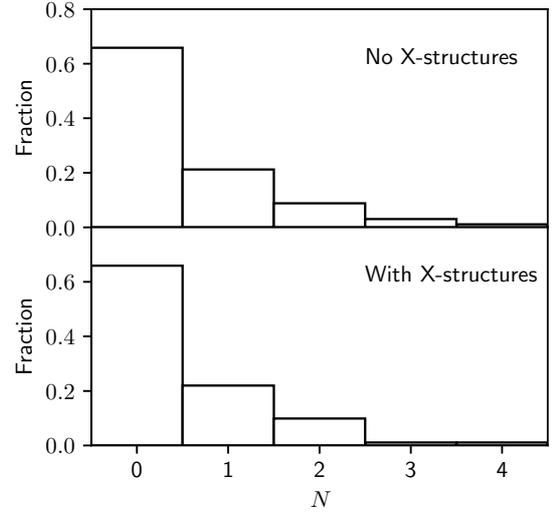}
    \caption{The distribution of the galaxies by the number of neighbours for
      the comparison sample (top) and our sample of galaxies with X-structures (bottom)
      according to the test 2 conditions (see text).}
    \label{fig:envirDistrib}
\end{figure}

Fig.~\ref{fig:envirDistrib} shows the distribution of galaxies of our sample
and the comparison sample by the number of neighbours according to the test 2
conditions.
The average number of neighbour galaxies for our sample is $0.24 \pm 0.05$
for the test 1 set and $0.54 \pm 0.10$ for the test 2  set (the standard 
errors of the mean are shown here). Corresponding values for the comparison samples
are $0.19\pm 0.05$ and $0.72\pm 0.14$. For both tests, the difference between the
main number of neighbours lies within $2\sigma$ limit, so we cannot consider this 
difference as statistically significant. We also used the two-sample Kolmogorov-Smirnov
test to check if the distributions of galaxies by the number of neighbours are 
statistically different. The $p$-value (the probability of the null hypothesis)
is 0.90 and 0.99 for the test 1 and the test 2 respectively, which means that
these distributions are statistically indistinguishable.

This result is in agreement with \citet{Shaw1987}, where the analysis
based on a sample of 23 galaxies with X-structures did not reveal a tendency 
of such galaxies to be located in a denser environment.

\subsection{The decomposition of galaxies}
\begin{figure*}
  \center
  \includegraphics[width=0.67\columnwidth, clip=true, trim=3.5mm 0mm 3.5mm 0mm]{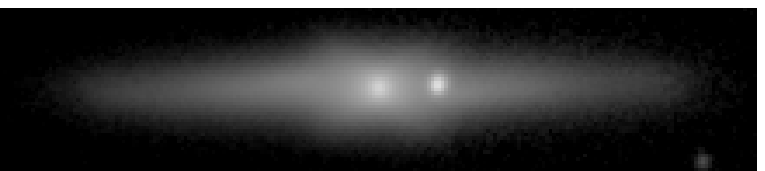} 
  \includegraphics[width=0.67\columnwidth, clip=true, trim=0mm 0mm 0mm 0mm]{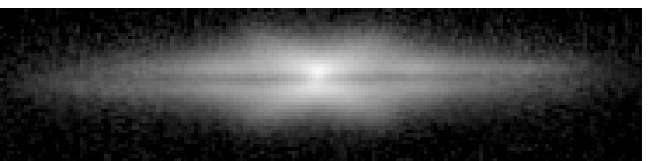}
  \includegraphics[width=0.67\columnwidth, clip=true, trim=0mm 4.3mm 0mm 4.3mm]{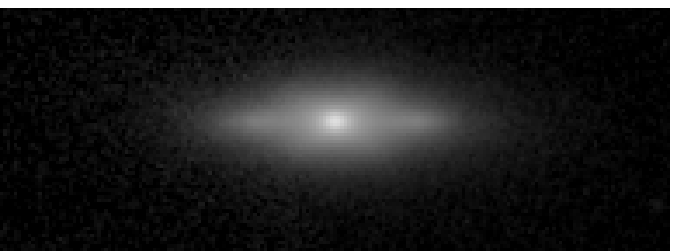}

  \includegraphics[width=0.67\columnwidth, clip=true, trim=3.5mm 0mm 3.5mm 0mm]{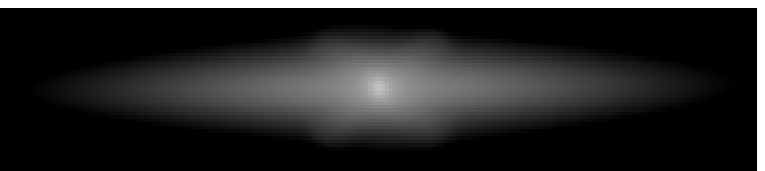}
  \includegraphics[width=0.67\columnwidth, clip=true, trim=0mm 0mm 0mm 0mm]{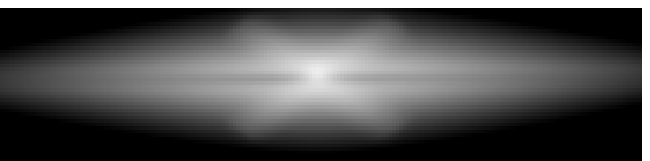}
  \includegraphics[width=0.67\columnwidth, clip=true, trim=0mm 4.3mm 0mm 4.3mm]{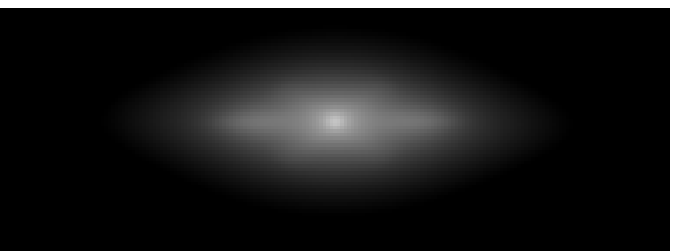}

  \includegraphics[width=0.67\columnwidth, clip=true, trim=3.5mm 0mm 3.5mm 0mm]{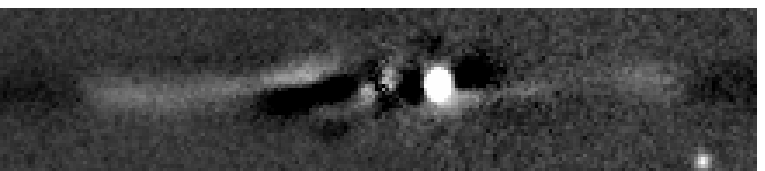}
  \includegraphics[width=0.67\columnwidth, clip=true, trim=0mm 0mm 0mm 0mm]{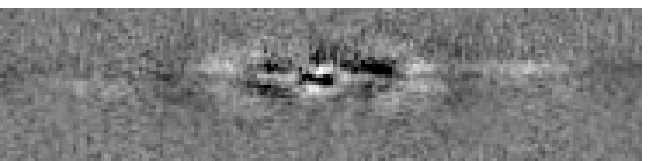}
  \includegraphics[width=0.67\columnwidth, clip=true, trim=0mm 4.3mm 0mm 4.3mm]{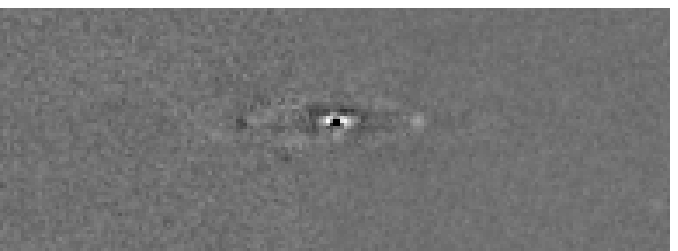}

  \caption{Results of the decomposition of three galaxies, 
  left to right: PGC~69401, PGC~53812, PGC~28788. Top panels --- 
  the SDSS $r$ band images, middle ones --- models, bottom ---
  residuals ``image$-$model''. Three different models are shown: 
  the most simple (bulge$+$disc$+$X-structure)
  for PGC~69401, with an additional dust lane for PGC~53812, and with 
  an additional edge-on ring for PGC~28788.}
  \label{fig:decompResults}
\end{figure*}

Fig.~\ref{fig:decompResults} demonstrates results of the 
decomposition for three galaxies: PGC~69401, PGC~53812, and PGC~28788. 
The top panels show original images of the galaxies in the $r$ band, the middle 
ones are the obtained models, and the bottom ones are the residuals (i.e., ``image--model'').
We applied three different models for these three galaxies. 
Our basic model consists of a disc, a bulge, and an X-structure 
(PGC~69401). The model of PGC~53812 contains the dust lane as an additional 
component, while the model of PGC~28788 also includes an edge-on ring (which can be 
seen as two bright blobs on both sides of the X-structure).

Below we provide results of the decomposition of galaxies 
from our subsample. We start with description of host galaxies 
(i.e., we list the parameters of main photometric components: the disc and the bulge) and then proceed with 
the parameters of the X-structure.

\subsubsection{Host galaxies}
Table~\ref{tab:decParams} shows main structural parameters of the host 
galaxies: 
the effective surface brightness $\mu_\mathrm{e}$, 
the effective radius $r_\mathrm{e}$, 
the S\'ersic index $n$ of the bulge; 
and the central surface brightness $\mu_0$, 
the radial $h$ and vertical $z_0$ scales, and the vertical shape of the light 
curve $m$ of the disc.
The values are given for the $r$ band for galaxies taken from SDSS and 
for the IRAC 3.6$\mu$m band for galaxies with available {\it Spitzer} observations 
(three bottom lines in the table); 
the absolute magnitudes, along with $\mu_\mathrm{e}$ and $\mu_0$ values, are corrected for 
the foreground Galaxy extinction according to the NED database \citep{Schlafly2011}.

The last column in Table~\ref{tab:decParams} shows the type of
additional components (a ring or a dust lane) which were used in the 
decomposition model. One can see that for most galaxies either a dust 
component or a ring was added. Moreover, almost all galaxies that do not 
have a prominent dust 
lane, contain a ring component. 
This feature can be common in galaxies with X-structures. 
In many cases the rings may be obscured by dust.

\begin{table*}
  \centering
  \begin{tabular}{lcccccccccc}
    \hline
    Name & $\mu_\mathrm{e} $ &  $r_\mathrm{e}$ &  $n$ &  $\mu_0$  & $h$&$m$& $z_0$&$l$  &  $\alpha$   & add. comp.\\
         & mag/$\square''$ &  kpc &      &  mag/$\square''$ &  kpc &     &  kpc & kpc ($h$)   &   deg  &  \\
    \hline
    PGC~24926 & 19.53 & 0.5 & 1.5 & 19.24 & 1.47 & 3.59  & 0.42 & 2.01 (1.37) & $21 \pm 3$ & dust \\
    PGC~26482 & 21.50 & 1.9 & 3.2 & 19.40 & 4.44 & 1.10  & 0.41 & 3.99 (0.90) & $28 \pm 2$ & dust \\
    PGC~28788 & 21.38 & 2.2 & 4.5 & 20.32 & 1.94 & 0.78  & 0.44 & 2.61 (1.35) & $26 \pm 1$ & ring \\
    PGC~28900 & 19.79 & 1.4 & 3.7 & 21.24 & 5.12 & 2.14  & 1.29 & 6.09 (1.19) & $22 \pm 1$ & ring \\
    PGC~32668 & 18.78 & 0.6 & 1.4 & 20.45 & 1.87 & 6.30  & 0.82 & 3.93 (2.10) & $29 \pm 0$ & ring \\
    PGC~37949 & 20.88 & 1.4 & 2.1 & 18.84 & 4.03 & 3.90  & 0.91 & 7.14 (1.77) & $28 \pm 4$ & dust \\
    PGC~39251 & 19.49 & 1.0 & 4.2 & 20.73 & 2.72 & 2.48  & 0.92 & 3.70 (1.36) & $15 \pm 2$ & ring \\
    PGC~44422 & 18.86 & 0.5 & 4.3 & 19.62 & 3.74 & 1.41  & 0.64 & 4.23 (1.13) & $23 \pm 2$ & ring \\
    PGC~69401 & 20.23 & 1.3 & 3.8 & 18.91 & 5.05 & 1.86  & 0.77 & 5.95 (1.18) & $27 \pm 3$ &  --- \\
    PGC~34913 & 21.44 & 1.3 & 2.5 & 18.74 & 4.34 & 20.00 & 0.66 & 3.56 (0.82) & $27 \pm 4$ & dust \\
    PGC~45214 & 19.81 & 1.5 & 0.7 & 17.93 & 3.72 & 12.91 & 0.59 & 7.74 (2.08) & $28 \pm 1$ & dust \\
    PGC~10019 & 20.62 & 2.1 & 1.2 & 18.37 & 3.65 & 6.18  & 0.54 & 5.88 (1.61) & $24 \pm 6$ & dust \\
    PGC~30221 & 20.87 & 1.6 & 1.6 & 19.08 & 3.09 & 2.98  & 0.47 & 4.32 (1.40) & $20 \pm 2$ & dust \\
    PGC~55959 & 21.05 & 1.9 & 2.3 & 18.41 & 3.94 & 20.00 & 0.25 & 6.02 (1.53) & $25 \pm 2$ & dust+ring\\
    ASK~361026.0&19.57 & 1.7 & 4.0 & 21.50 & 5.79 & 1.30  & 1.38 & 6.83 (1.18) & $22 \pm 2$ & ring \\
    PGC~53812 & 21.00 & 0.9 & 1.0 & 19.10 & 5.42 & 10.94 & 1.08 & 7.10 (1.31) & $26 \pm 1$ & dust \\
    PGC~02865 & 24.25 & 4.2 & 3.5 & 17.50 & 3.28 & 3.46  & 0.41 & 4.10 (1.25) & $32 \pm 1$ & dust \\
    PGC~21357 & 20.77 & 4.5 & 0.8 & 19.76 & 5.54 & 4.24  & 0.40 & 4.54 (0.82) & $35 \pm 3$ & dust \\
    PGC~69739 & 22.31 & 2.6 & 4.0 & 19.66 & 5.88 & 1.66  & 0.61 & 3.76 (0.64) & $27 \pm 1$ & dust \\
    \hline
    IC~2531    & 21.11 & 1.2 & 4.7 & 24.07 & 8.25 & 20.00 & 0.62 & 2.81 (0.34) & $29\pm 1$ &  --- \\
    NGC~4013   & 17.63 & (0.1) & 1.8 & 22.52 & 1.90 & 20.00 & 0.38 & 0.68 (0.36) & $22\pm 1$ &  --- \\
    NGC~5529   & 21.79 & 1.2 & 3.3 & 22.84 & 8.00 & 20.00 & 0.25 & 2.00 (0.25) & $38\pm 2$ &  --- \\
    \hline
  \end{tabular}
  \caption{
  The structural parameters of the subsample galaxies: the effective surface brightness ($\mu_\mathrm{e}$), the effective
  radius ($r_\mathrm{e}$) and the S\'{e}rsic index ($n$) of the bulge; the central surface brightness ($\mu_0$),
  the radial exponential scale ($h$), the $m$-value and the vertical scale ($z_0$) of the disc;
  the mean length of X-structure rays ($l$) and the mean angle between rays and
  the disc plane ($\alpha$). All values are estimated in the $r$ band. The last column shows the
  additional components of the model.
  The three bottom rows contain characteristics of the galaxies observed with the
    {\it Spitzer} telescope.}
  \label{tab:decParams}
\end{table*}

We should emphasize that the values provided in Table~\ref{tab:decParams} 
may be distorted by the dust lane attenuation because all galaxies 
are observed at the edge-on orientation where the dust influence is  
severe. This is especially significant for the bulges, which
are seen entirely inside of the dust lane. However, our main 
goal of this decomposition is to separate the host galaxy light from 
the light of the X-structure, such that it can be analysed independently.

The mean value of the exponential scale 
in our subsample is $\langle h \rangle = 3.9 \pm 1.3$~kpc
(only galaxies from SDSS are taken into account). This means that the galaxies in
our sample are comparable to or larger than the Milky Way \citep[the exponential scale lengths
of the Milky Way are $3.00 \pm 0.22$ and $3.29 \pm 0.56$ kpc for the thin and thick
disks respectively according to][]{McMillan2011}.

\subsubsection{The X-structures}
\begin{figure*}
  \center
  \includegraphics[width=0.4\columnwidth, clip=true, trim=2mm 0mm 3mm 0mm]{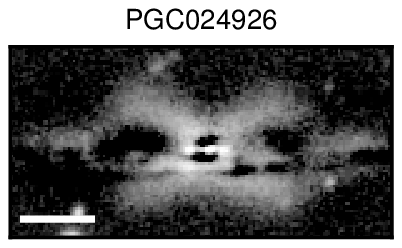}
  \includegraphics[width=0.4\columnwidth, clip=true, trim=2mm 0mm 3mm 0mm]{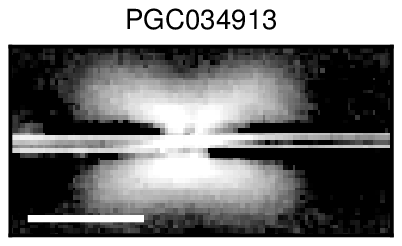}
  \includegraphics[width=0.4\columnwidth, clip=true, trim=2mm 0mm 3mm 0mm]{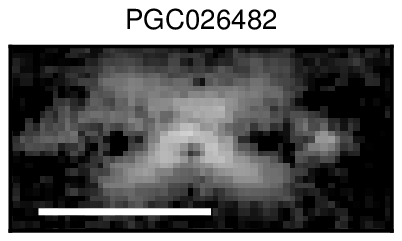}
  \includegraphics[width=0.4\columnwidth, clip=true, trim=2mm 0mm 3mm 0mm]{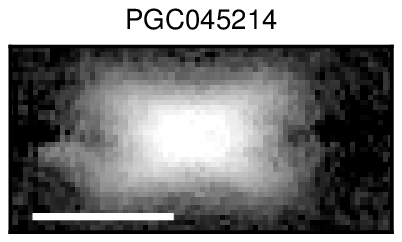}
  \includegraphics[width=0.4\columnwidth, clip=true, trim=2mm 0mm 3mm 0mm]{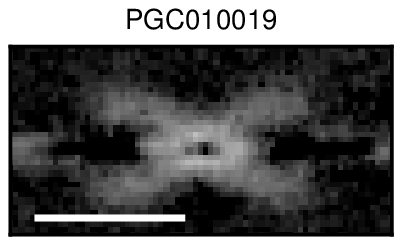}
  \includegraphics[width=0.4\columnwidth, clip=true, trim=2mm 0mm 3mm 0mm]{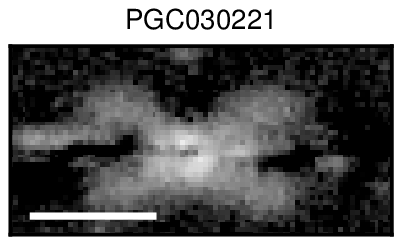}
  \includegraphics[width=0.4\columnwidth, clip=true, trim=2mm 0mm 3mm 0mm]{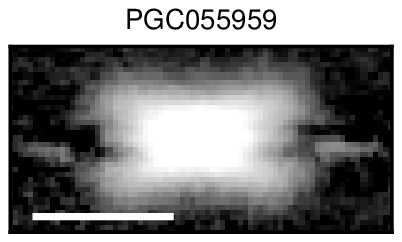}
  \includegraphics[width=0.4\columnwidth, clip=true, trim=2mm 0mm 3mm 0mm]{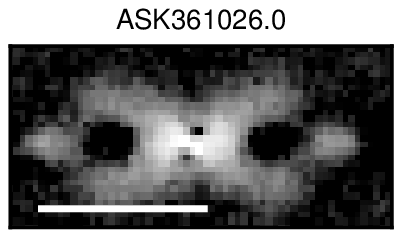}
  \includegraphics[width=0.4\columnwidth, clip=true, trim=2mm 0mm 3mm 0mm]{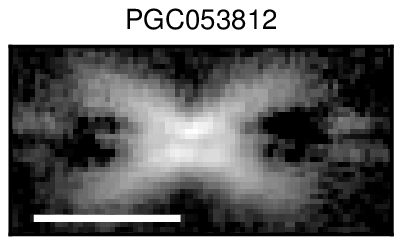}
  \includegraphics[width=0.4\columnwidth, clip=true, trim=2mm 0mm 3mm 0mm]{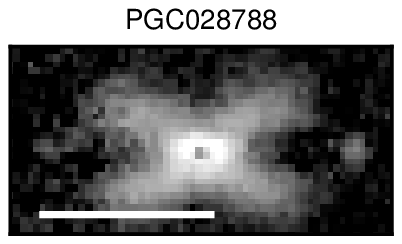}
  \includegraphics[width=0.4\columnwidth, clip=true, trim=2mm 0mm 3mm 0mm]{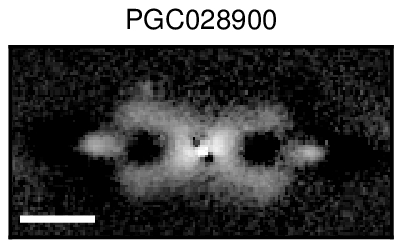}
  \includegraphics[width=0.4\columnwidth, clip=true, trim=2mm 0mm 3mm 0mm]{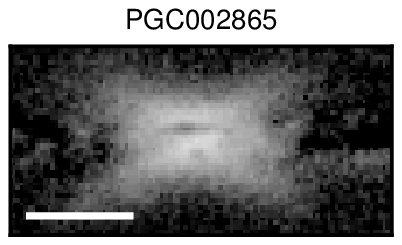}
  \includegraphics[width=0.4\columnwidth, clip=true, trim=2mm 0mm 3mm 0mm]{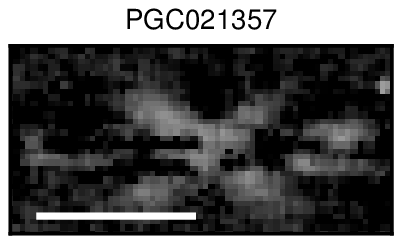}
  \includegraphics[width=0.4\columnwidth, clip=true, trim=2mm 0mm 3mm 0mm]{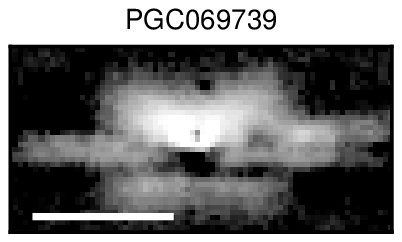}
  \includegraphics[width=0.4\columnwidth, clip=true, trim=2mm 0mm 3mm 0mm]{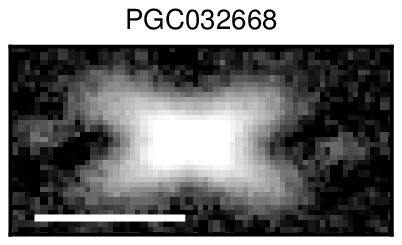}
  \includegraphics[width=0.4\columnwidth, clip=true, trim=2mm 0mm 3mm 0mm]{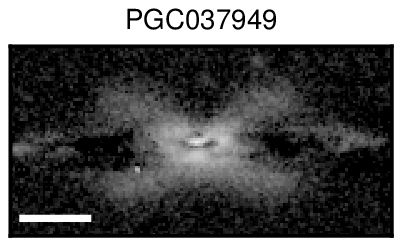}
  \includegraphics[width=0.4\columnwidth, clip=true, trim=2mm 0mm 3mm 0mm]{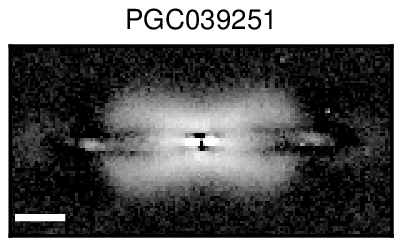}
  \includegraphics[width=0.4\columnwidth, clip=true, trim=2mm 0mm 3mm 0mm]{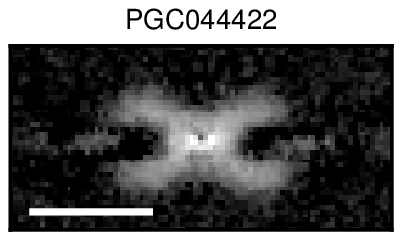}
  \includegraphics[width=0.4\columnwidth, clip=true, trim=2mm 0mm 3mm 0mm]{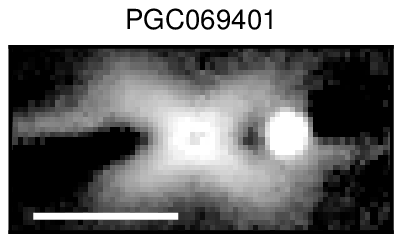}\\
  \includegraphics[width=0.4\columnwidth, clip=true, trim=2mm 0mm 3mm 0mm]{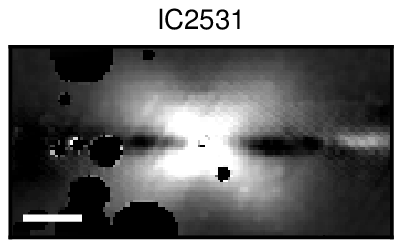}
  \includegraphics[width=0.4\columnwidth, clip=true, trim=2mm 0mm 3mm 0mm]{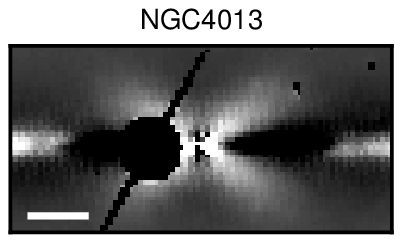}
  \includegraphics[width=0.4\columnwidth, clip=true, trim=2mm 0mm 3mm 0mm]{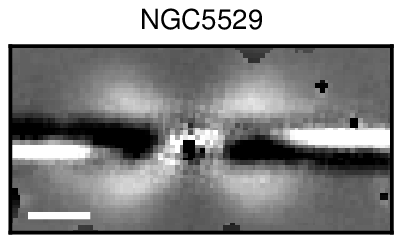}
  \caption{The central regions of the subsample galaxies after the host galaxy subtraction. 
  The central X-structures are clearly visible. IRAC 3.6$\mu$m images for the three galaxies
  are shown in the bottom row. The white bars in the bottom left corner of each image show the 10'' scale. 
  The dark features in the images are masked 
  regions.}
  \label{fig:xOnly}
\end{figure*}

Fig.~\ref{fig:xOnly} shows central regions of the images after 
subtracting from them the obtained models for the host galaxies.
Different kinds of the
X-structure shapes are clearly
distinguishable: almost box-like in PGC~45214, with straight ``rays'' in PGC~28788 and PGC~32668,
and the infinity sign-shaped in PGC~28900. 
One can also notice different axis ratios of the X-structures.  

\begin{figure}
  \center
  \includegraphics[width=0.9\columnwidth, clip=true, trim=0mm 0mm 0mm 0mm]{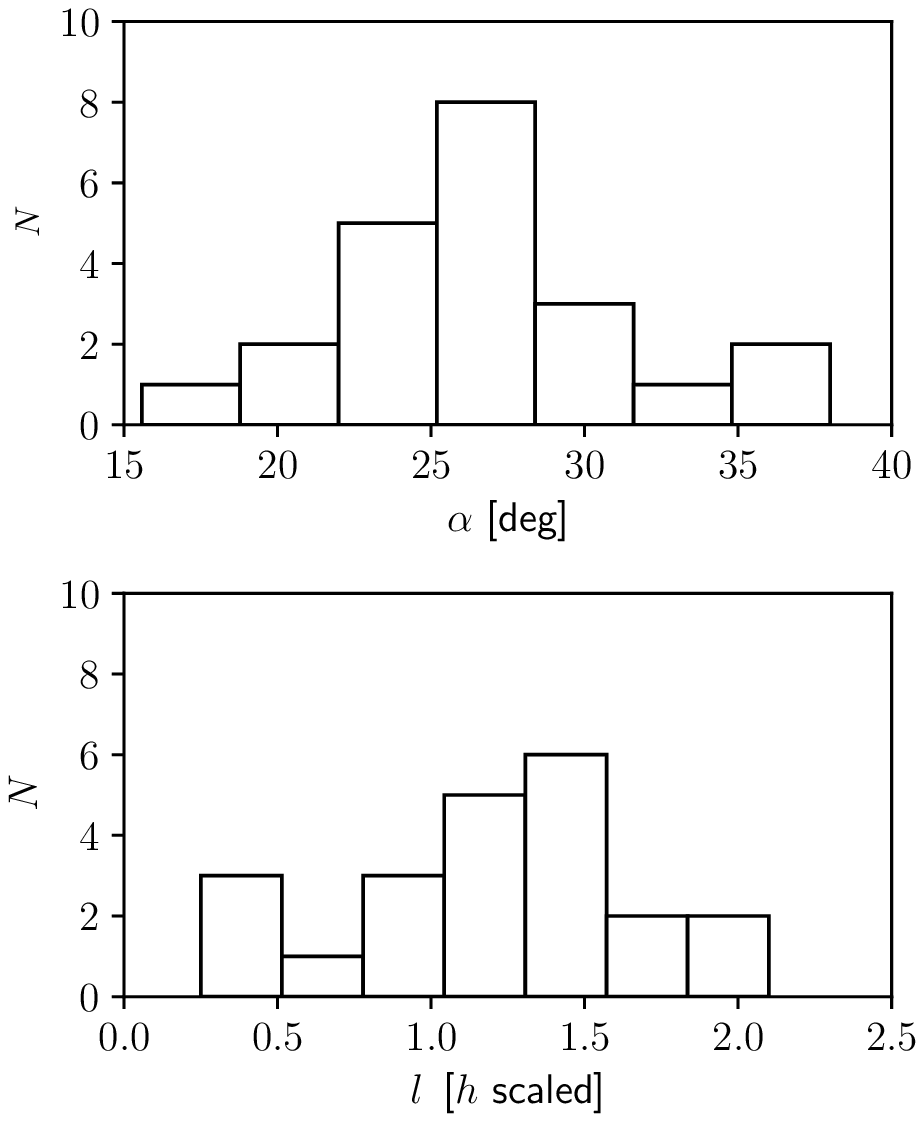}
  \caption{The distribution of the angle $\alpha$ (top) and
  lengths normalized by $h$ (bottom) of the 
  X-structures.}
  \label{fig:xHist}
\end{figure}

In the last two columns of Table~\ref{tab:decParams} we list the
parameters of the X-structures:
the mean length of the rays $l$ measured in kpc and normalised by $h$ 
(values in brackets), and the mean angle $\alpha$ between the rays and 
the disc midplane. Fig.~\ref{fig:xHist} demonstrates the distributions of the
$\alpha$ and $l$ in the 
X-structures. 
The average values of these parameters are
$\langle \alpha \rangle = (26.1 \pm 5.1)$\degr and $\langle l \rangle = (1.18 \pm 0.50) h$.

Since the most favoured scenario of the X-structure formation includes 
a bar, we decided to compare the observed distribution of the sizes of the 
X-structures for galaxies of our subsample with the distribution of bar 
sizes for galaxies viewed at a non edge-on orientation (i.e. for
galaxies whose bars can be directly observed). We took observed 
bar sizes from \citet{Gadotti2009}, where the decomposition results of 
$\sim 1000$ SDSS galaxies are provided. The mean exponential scale
of barred galaxies from this sample is $3.20 \pm 1.25$ kpc, which is similar to
the mean value of our sample $3.95 \pm 1.30$ kpc.

\begin{figure}
  \center
  \includegraphics[width=\columnwidth]{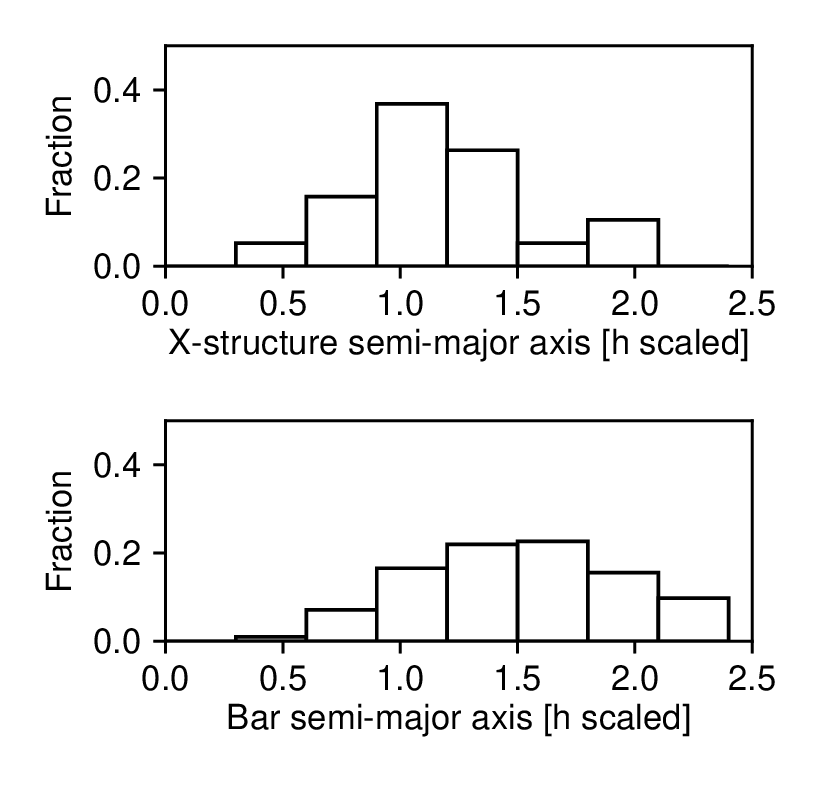}
  \caption{The size of the X-structures obtained in this work (top)
    versus the bar size from \citet{Gadotti2009} (bottom). Values are
    normalised to the disk exponential scale-length.}
  \label{fig:barCompar}
\end{figure}

Fig.~\ref{fig:barCompar} shows 
the comparison of the X-structure semi-major axis size distribution 
(in projection on the galactic midplane) and the distribution of the bar sizes, 
both scaled to the disc exponential scale. One can see that the length
of the semi-major axis for the most extended bars is approximately $2.5\,h$,
therefore, one should not expect the X-structures to be
larger than that. Indeed, The semi-major axis of all X-structures
in our sample lies below this value. Moreover, the size of the X-structures
is systematically smaller than the size of the bars: the maximum of the
distribution of the X-structure sizes lies just above 1 kpc, whereas
the distribution of the bar sizes reaches its maximum at about 1.7 kpc. This result
is consistent with previous conclusions by \citet{Erwin2013},
\citet{HerreraEndoqui2015}, and \citet{Erwin2017}, who found that
the spatial extent of the X-structures is about 0.38--0.54 of the bar length
in moderately inclined galaxies.

\begin{figure}
  \center
  \includegraphics[width=\columnwidth]{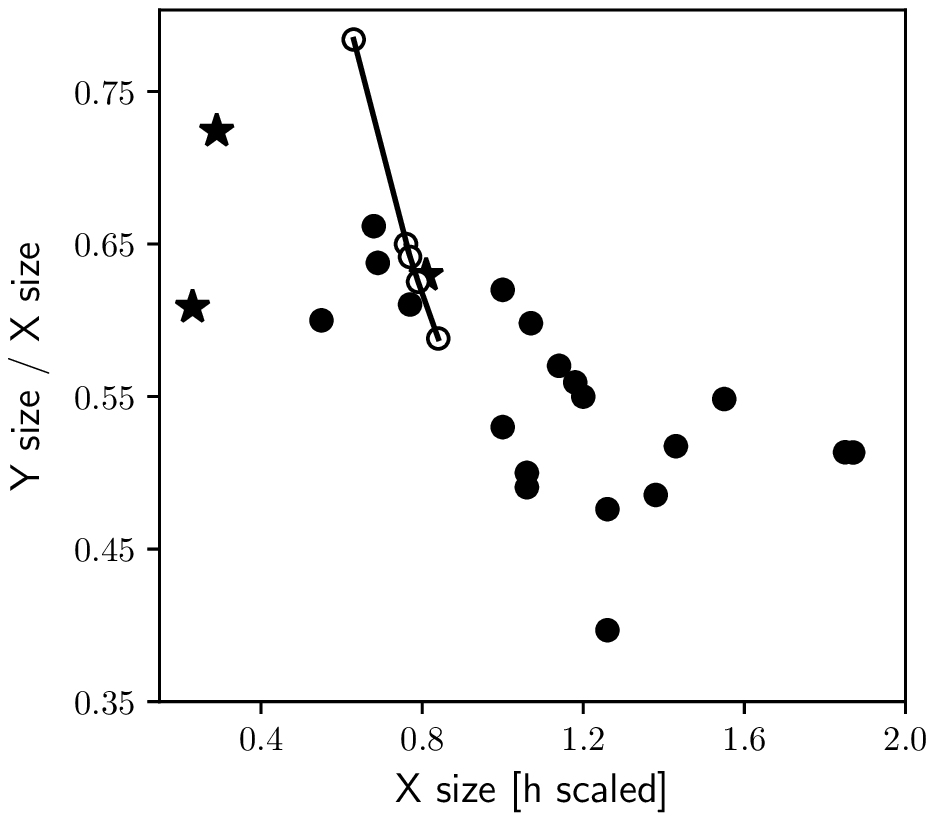}
  \caption{The relation between the X-structure semi-major axis size and
    its minor-to-major axis ratio. Circles --- SDSS objects, 
    stars --- objects observed in the IRAC 3.6$\mu$m band. The open circles connected 
    with the line represent the $N$-body simulations with different orientation (PA).
    (see Seq.~\ref{seq:nbody_decomp}).}
  \label{fig:xSizeQ}
\end{figure}

Fig.~\ref{fig:xSizeQ} shows the relationship between the observed 
size of the X-structures and their vertical-to-lateral axis ratio. 
One can see that bigger X-structures have smaller axis ratios, in 
general. This can be easily explained within the
bar-driven scenario. The observed length of the X-structure 
depends on the orientation of the bar with respect to the line of 
sight: the length is the largest when the bar is oriented perpendicular
to the line of sight and gradually decreases if the bar is rotated toward
the end-on orientation. 
On the other hand, the observed height of the X-structure should remain the same regardless
of the bar orientation (as long as the observer stays within the disc 
midplane). Therefore, the observed height-to-length ratio varies with changing of the bar viewing angle.

\subsubsection{The case of NGC~4013 and NGC~5529}

As was mentioned above, two galaxies in our subsample, NGC~4013 and NGC~5529, have
both SDSS and {\it Spitzer} images. 

The 
dust lanes in these galaxies are extremely prominent and our 
approximate method of taking dust into account by adding a new dust
component failed to  fit a model. Instead, we started with 
decomposition in the IRAC 3.6$\mu$m band where the dust influence is relatively
small. Then we masked out dusty regions in the optical images and used 
the IR model as a starting point for the decomposition in the optical bands.
The relatively wide PSF of the 3.6$\mu$m image did not allow us to resolve a bulge
for NGC~4013 and, therefore, the decomposition process converged to an unrealistic 
model with $r_\mathrm{e}=0.1$~kpc.

Unfortunately, the inner parts of the X-structures in the optical bands are 
completely hidden by the dust lanes in these two galaxies. This 
circumstance did not allow us to perform the described above
X-structure analysis and to obtain $\alpha$ and $l$ values in order to compare 
them with their IR counterparts. The only reliable
parameter that can be compared 
is the ellipticity $e=1-b/a$ of the X-structure obtained from the 
photometric model. This parameter has close values for the optical 
($r$ band) and IR images: 0.32 and 0.28 for NGC~4013, 0.34 and 0.34 
for NGC~5529.

\section{N-body modelling}
\label{sec:Nbody}
\subsection{Simulations}
In order to interpret the results of our photometric decomposition, we 
perform a set of $N$-body simulations and use snapshots of
obtained models as input data to the procedure that we applied
to the subsample of real galaxies. 
In our study we adopted a fiducial $N$-body model of the Milky Way which has been considered by ~\citet{McMillan2007}. 

The initial conditions of our simulation comprise three components: 
the disc, the bulge, and the dark halo. 
All three are live, i.e. described self-consistently, 
in order to allow exchange of angular momentum and, thus, a full bar 
growth 
\citep{Athan2002,Athan2003}. 

The initial disc has an exponential surface density profile 
with a vertical structure modeled with isothermal sheets
\begin{equation}
 \rho_\mathrm{d} (r, z) = 
 \frac {M_\mathrm{d}}{4 \pi h^2 z_0} \, 
 \exp(-r/h) \, \mathrm{sech}^2 
 \bigg(\frac{z}{z_0}\bigg),
\end{equation}
where $R$ is the cylindrical radius, 
$h$ is the disc radial scale length, 
$z_0$ is the disc scaleheight, and 
$M_\mathrm{d}$ is the total disc mass. 

The dark halo model is a truncated NFW halo~\citep{NFW1996,NFW1997}
with the initial volume density
\begin{equation} 
\rho_\mathrm{h} (r) =
\frac{C_\mathrm{h} \cdot T(r/r_\mathrm{t})}
{(r/r_\mathrm{h})(1 + r/r_\mathrm{h})^{2}} \, ,
\label{eq:tNFW} 
\end{equation}
with $T(r/r_\mathrm{t}) = 2/(\exp{(r/r_\mathrm{t})} + \exp(-r/r_\mathrm{t}))$, 
where $r$ is the radius, $r_\mathrm{h}$ and $r_\mathrm{t}$ are the halo core 
and cut-off radii, respectively, 
and the constant $C_\mathrm{h}$ is a parameter defining the full mass of 
the halo $M_\mathrm{h}$. 

The bulge is modeled with a \cite{Hernquist1990} density profile
\begin{equation} 
\label{eq:rhohern}
\rho_\mathrm{b} (r) = 
\frac{M_\mathrm{b} \, r_\mathrm{b}}
{2 \pi r (r_\mathrm{b} + r)^{3}} \, ,
\end{equation}
where  
$r_\mathrm{b}$ is a scale parameter and 
$M_\mathrm{b}$ is the total bulge mass.

We choose the units of these parameters such that the Newton's constant of gravity $G = 1$,
$h = 1$, and $M_\mathrm{d} = 1$. 
We  assume that the initial disc scale height is $z_0 = 0.05$, 
the bulge mass is $M_\mathrm{b} = 0.2$, and scale length 
is $r_\mathrm{b} = 0.2$. 
Scaling these values to the Milky Way disc, taking $h = 3.5\,$kpc and
$M_\mathrm{d} = 5\times10^{10}\,\mathrm{M}_\odot$, gives 
the time unit of $\simeq 14\,$Myr, and
the velocity unit of $\sim 250\,$km$\,$s$^{-1}$.

For the halo we take the scale radius as in~\cite{McMillan2007} 
$r_\mathrm{h} = 6$, the truncation radius $r_\mathrm{t} = 60$, and the mass 
$M_\mathrm{h} = 24$. Within a sphere of radius $r = 4h$ the ratio 
$M_\mathrm{h} (4h)/M_\mathrm{d} = 1.56$.

In our tests the halo has $1\,200\,000$ particles, 
the bulge -- $40\,000$ and the disc -- $200\,000$. 

The initial conditions are built using the script 
{\tt mkgalaxy}~\citep{McMillan2007} from the package 
NEMO~\citep{Teuben1995}. In the first step, the spherical initial 
conditions for the halo and the bulge in the presence of the
monopole part of the disc potential as external potential 
are generated. Then both spheroids are adjusted to the presence 
of the full disc potential, rather than to only its monopole. The 
final step is to populate the disc component. In this step 
the potential of the halo and bulge are 
considered as external. The smoothed, azimuthally averaged potential 
of these components is described by the potential expansion.

The model evolves from the initial conditions with a dynamically 
cold disc, such that the Toomre parameter \citep{Toomre1964} 
$Q = 1.2$ at radius $R=2h$. To simulate the evolution of 
the models, we use the fast $N$-body code {\tt gyrfalcon} 
\citep{Dehnen2000,Dehnen2002}. The gravitational softening lengths 
are $\epsilon = 0.02$ for the disc and bulge particles, and $0.04$ the for 
halo particles. We vary the integration time step according to the 
rule $0.1 \sqrt{\epsilon/|\vec a|}$, where $\vec a$ is the 
gravitational acceleration. The opening angle was set to 0.6. 

\begin{figure}
  \center
  \includegraphics[width=1.0\columnwidth]{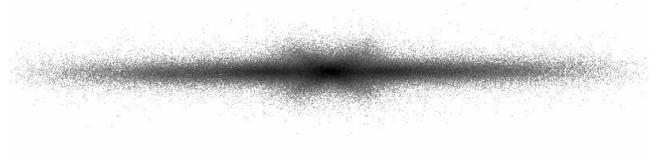}
    \caption{Log-scale density diagram of the edge-on view along the bar
      minor axis of the simulated disc. The presence of an X-shaped
      structure is clearly visible.}
    \label{fig:nbodyModel}
\end{figure}

It has been long recognized that cool discs are unstable with respect 
to the bar-like instability (see a review by \citealp{Sellwood2013}). 
In both cases we observe the formation of a 
bar after $t=50$. The bar grows, and after $t=100$ it experiences the 
buckling instability that results in the formation of an X-structure. 
Fig.\ref{fig:nbodyModel} shows the disc at $t=200$. 

\subsection{The analysis of the $N$-body simulated images}
\label{seq:nbody_decomp}
In this section we present results of the decomposition of 
our $N$-body model images described above. The big advantage of a 
$N$-body modelling approach is that we can control the orientation of the galaxy 
and investigate how the observed parameters of the X-structure vary when we change the point of view. 

We performed the decomposition of simulated $N$-body model images generated  for
different points of view. We took $N$-body snapshots for t = 200 at different
points of view and gridded them into a 2D dataset by binning the $x$-$z$ coordinates
of all particles to a regular rectangular grid, keeping in each cell the logarithm
of the number of particles. Then we used the {\tt NEMO} tools to convert the gridded
image into the FITS-format. The prepared this way data we considered as 'observables'
and analyzed them as described in Sect.~\ref{sec:methods}.

\begin{figure*}
  \center
  \includegraphics[width=0.6\columnwidth, clip=true, trim=0 0 1mm 1mm]{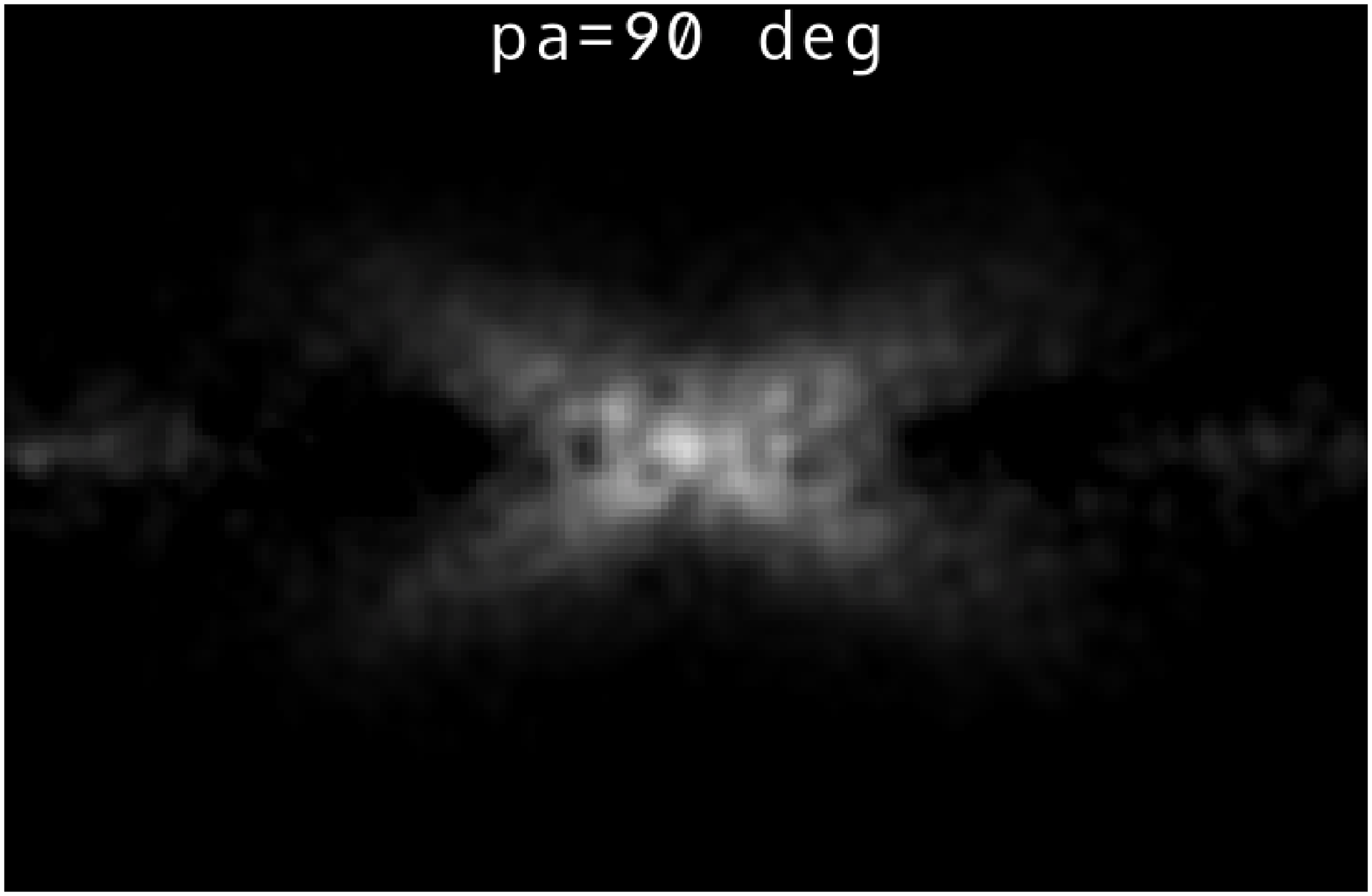}
  \includegraphics[width=0.6\columnwidth, clip=true, trim=0 0 1mm 1mm]{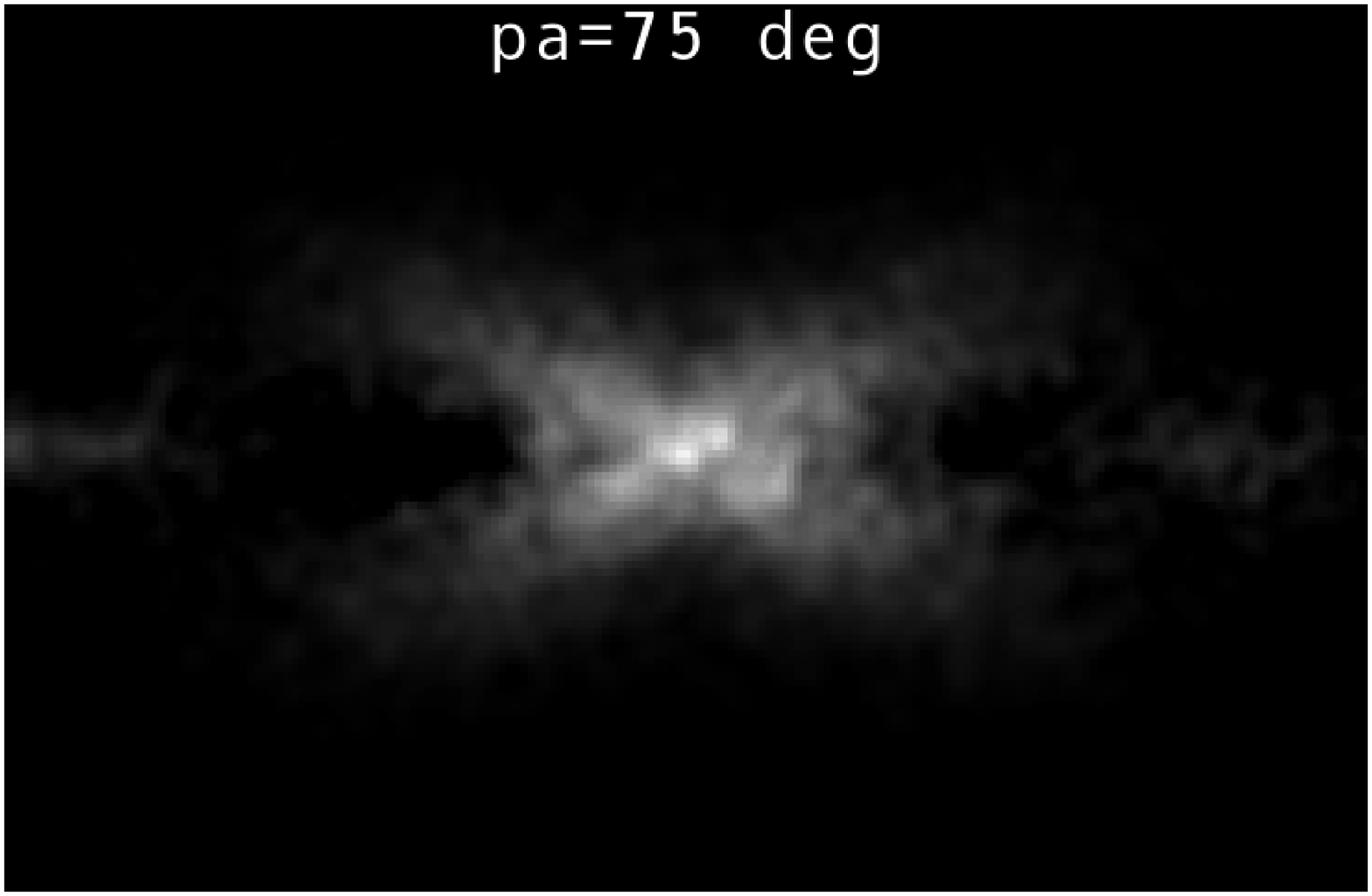}
  \includegraphics[width=0.6\columnwidth, clip=true, trim=0 0 1mm 1mm]{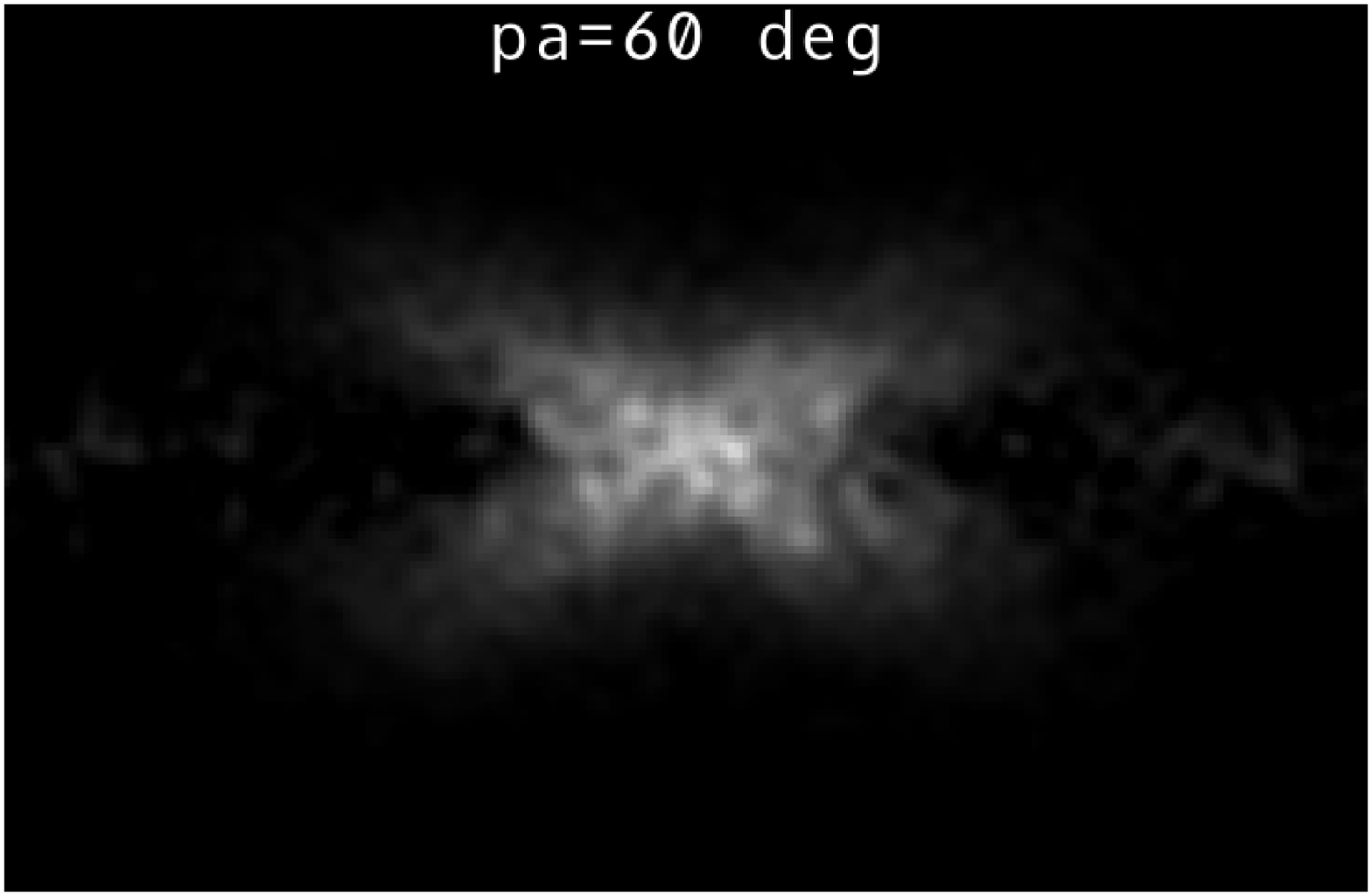}
  \includegraphics[width=0.6\columnwidth, clip=true, trim=0 0 1mm 1mm]{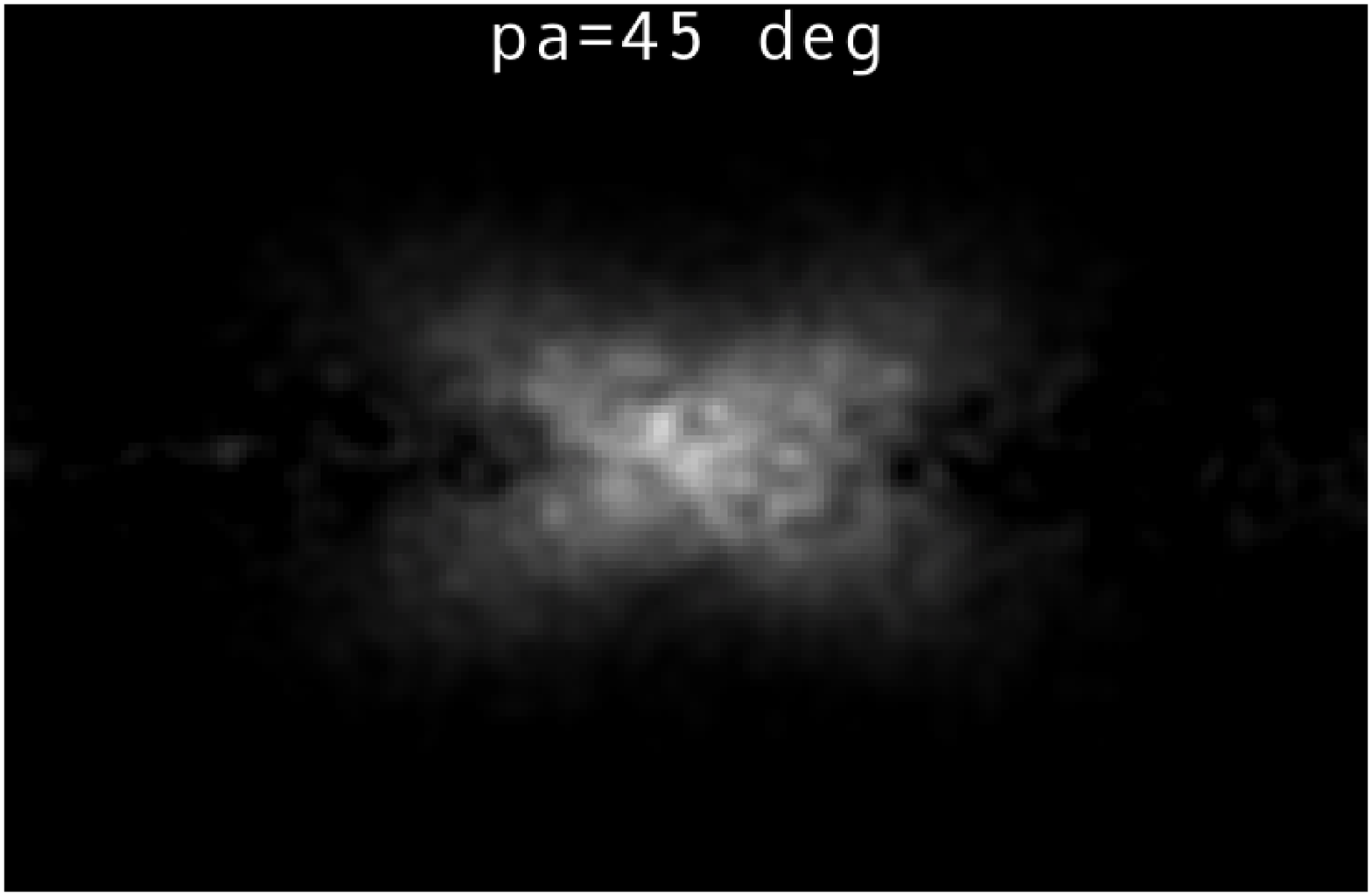}
  \includegraphics[width=0.6\columnwidth, clip=true, trim=0 0 1mm 1mm]{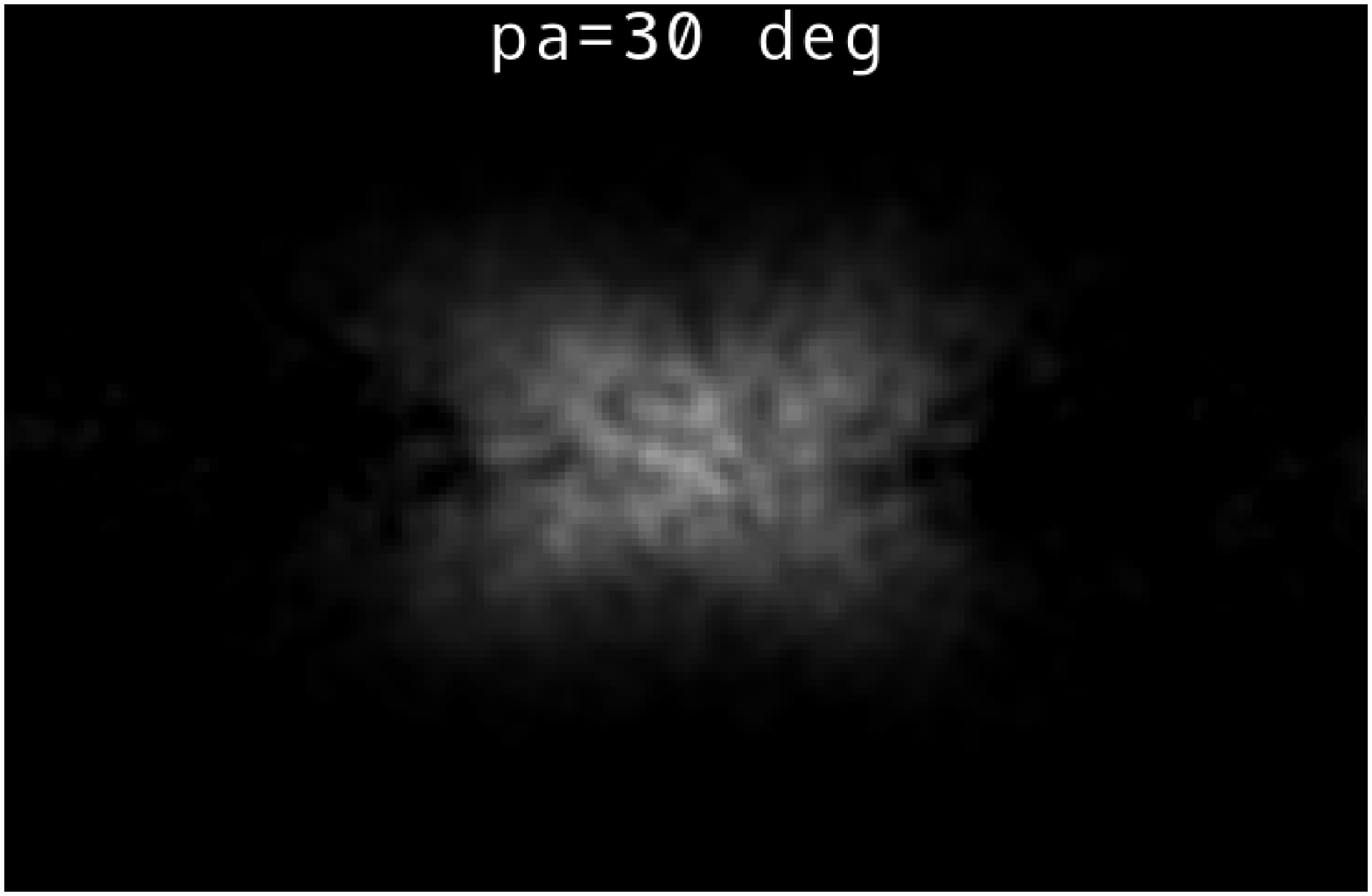}
  \includegraphics[width=0.6\columnwidth, clip=true, trim=0 0 1mm 1mm]{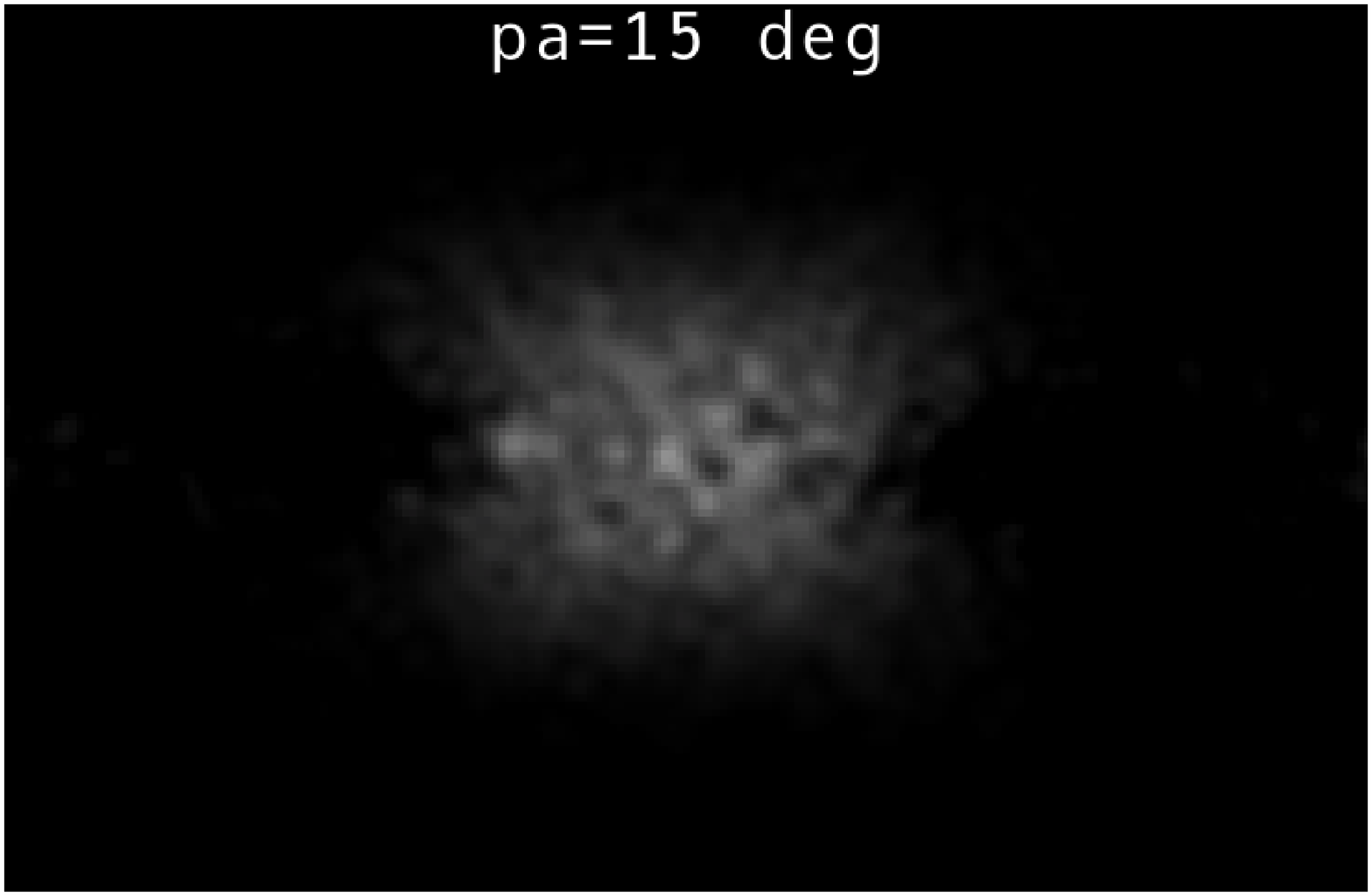}
  \caption{The images of the $N$-body models after the subtraction of the host 
  model for different viewing angles from 90\degr 
  (the bar is perpendicular to the line of sight) to 15\degr 
  (the bar is almost parallel to the line of sight). 
  The model images were convolved with a Gaussian kernel with FWHM of 
  3 pixels.
  }
  \label{fig:nBodyX}
\end{figure*}

Fig.~\ref{fig:nBodyX} shows a set 
of $N$-body images after the subtraction of 
the host photometric model. These images represent the same snapshot 
of the $N$-body model ($t=200$) viewed from different
angles: the bar position angle varies from 90\degr (the bar is perpendicular 
to the line of sight) and to 15\degr (the bar is almost parallel to the 
line of sight). One can clearly observe how appearance of the
X-shaped structure gradually changes from a strong X-like shape to a
boxy one with the rotation of the galaxy.

\begin{figure}
  \center
  \includegraphics[width=0.8\columnwidth, clip=true, trim=0 0 0mm 0mm]{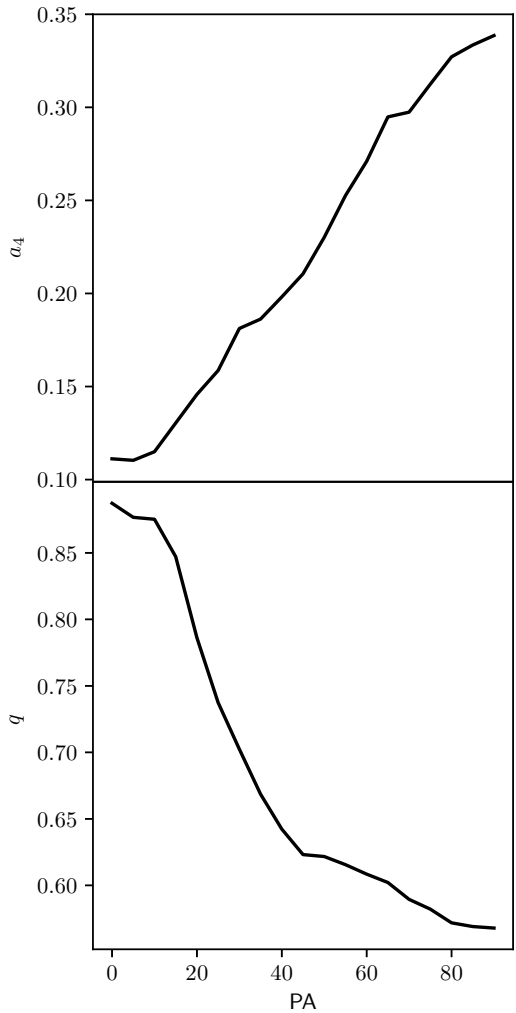}
  \caption{The dependence of the $a_4$ Fourier 
  amplitude (top) and the axis ratio 
  (bottom) of the X-structure as a function of the bar position angle. 
  $\mathrm{PA}=0\degr$ 
  is the orientation when the bar is parallel to the line of sight.}
  \label{fig:nBodyGraphs}
\end{figure}

Fig.~\ref{fig:nBodyGraphs} demonstrates how the $a_4$ Fourier mode and 
the X-structure axis ratio changes with the bar orientation. 
As expected, the Fourier mode grows gradually as the bar
orientation goes toward the side-on orientation, while the axis ratio gradually 
decreases. We argue that the behaviour of the
axis ratio, which is shown in the bottom panel of Fig.~\ref{fig:nBodyGraphs}, 
is responsible for the relation shown in Fig.~\ref{fig:xSizeQ}, 
and which we observe in the real galaxies of our subsample.

To show how the galaxy moves in Fig.~\ref{fig:xSizeQ} with changing
its position angle, we measured the size of the $N$-body X-structure
along the $x$- and $y$-axes for different orientations of the model and 
ploted these values, along with the data for the subsample galaxies (white circles).
The rightmost point corresponds to the bar orientation perpendicular 
to the line of sight ($0\degr$), and each next point 
demonstrates the
addition of 15 degrees, i.e., the leftmost point shows the $60\degr$ 
orientation. One can see how the galaxy moves to the left when 
the position angle increases. A high value of the minor-to-major axis ratio 
for the leftmost point, which lies considerably above the points for the subsample 
galaxies, can be explained by a high value of the position angle for 
this point: real galaxies viewed from this angle probably will not 
demonstrate a prominent X-structure.

To compare the observed size of the X-structure with the bar size, we gridded the face-on projection of our $N$-body snapshot.
Then we made a contour representation of a 2D image map and defined the 
distance along the major axis of a bar where the ellipticity of isodensities drops abruptly
to zero. This distance estimates the extension of a bar 1.75$h$. The procedure is similar to
that described by \citet{Athanassoula2002}. The size of the X-structure is about a
half of the bar size. This result is in agreement with previous findings.

Another result we  notice is the difference between the axis ratio
in the real and simulated galaxies. The axis ratio 
in the models spans roughly from 0.6 to 0.8 for different orientations, whereas the
axis ratio in the real galaxies can be as low as 0.4. Our model does not
explain the existence of very flattened X-structures with the axis ratio
less than $\sim 0.6$. Perhaps, such flattened structures may form 
in galaxies with more massive dark haloes,
 but it is a subject of a further investigation.

\section{Conclusions}

In this work we consider a sample of galaxies with prominent 
X-shaped structures at their centres. We develop a procedure
for estimating the geometric parameters of the X-structures along 
with the parameters of their host galaxies. The procedure is
applied to the subsample of galaxies.
In order to interpret the obtained results, we perform $N$-body simulations 
of a Milky Way-like galaxy and apply the same decomposition 
procedure to the  simulated images. 

Our main results are as follows:
\begin{enumerate}
\item galaxies with the X-structures reside in approximately the same
local spatial environment as galaxies without such features;

\item we estimate the
structural parameters of the X-structures;

\item the distribution of the X-structures by its apparent size shows a maximum at
about 1.1 disc exponential scale lengths which is in agreement with
previous findings drawn by other authors for face-on galaxies.

\item There is a relation between the observed size of the X-structure and 
its axis ratio, which can be explained by the projection effect 
when a bar is viewed from different position angles.

\item We applied our decomposition procedure to the images created via the $N$-body modelling. 
The results for them are in qualitative agreement with those for the real galaxies.
\end{enumerate}

Our main results confirm the bar-driven scenario of the X-structure 
origin. Although the analysis of our $N$-body simulations is in agreement with the results for the
subsample galaxies, the characteristics of the X-structure for some galaxies significantly differ from the simulated ones. 
This requires a further investigation on a set of realistic $N$-body models with different input parameters,
including the shape and the mass of the dark halo.

Another direction of a further development of this work is to apply our
approach to a wider sample of galaxies. Application of our methods to the
subsample, based on the optical SDSS images, showed that obscured by dust central regions
of a galaxy is a big issue. A possible solution is to use
edge-on galaxies with prominent X-structures from the 
S4G\footnote{http://irsa.ipac.caltech.edu/data/SPITZER/S4G/} catalogue \citep{Sheth2010}. 
The fact that we have successfully applied our procedure to the NIR images of the three
galaxies, for which it failed in optics, shows that it is
possible to apply 
our
method for a larger sample of objects.

\section*{Acknowledgements}

We thank the anonymous reviewer for a thorough review and highly appreciate the comments and 
suggestions, which allowed us to significantly improve the quality of the publication.

Aleksandr Mosenkov is a beneficiary of a postdoctoral grant from the
Belgian Federal Science Policy Office. Aleksandr Mosenkov also acknowledges support from
St. Petersburg University research grant 6.38.335.2015.

DB acknowledges support from RSF grant RSCF-14-50-00043.

Funding for the Sloan Digital Sky Survey IV has been provided by
the Alfred P. Sloan Foundation, the U.S. Department of Energy Office of
Science, and the Participating Institutions. SDSS-IV acknowledges
support and resources from the Center for High-Performance Computing at
the University of Utah. The SDSS web site is www.sdss.org.

SDSS-IV is managed by the Astrophysical Research Consortium for the 
Participating Institutions of the SDSS Collaboration including the 
Brazilian Participation Group, the Carnegie Institution for Science, 
Carnegie Mellon University, the Chilean Participation Group, the French
Participation Group, Harvard-Smithsonian Center for Astrophysics, 
Instituto de Astrof\'isica de Canarias, The Johns Hopkins University, 
Kavli Institute for the Physics and Mathematics of the Universe (IPMU) / 
University of Tokyo, Lawrence Berkeley National Laboratory, 
Leibniz Institut f\"ur Astrophysik Potsdam (AIP),  
Max-Planck-Institut f\"ur Astronomie (MPIA Heidelberg), 
Max-Planck-Institut f\"ur Astrophysik (MPA Garching), 
Max-Planck-Institut f\"ur Extraterrestrische Physik (MPE), 
National Astronomical Observatory of China, New Mexico State University, 
New York University, University of Notre Dame, 
Observat\'ario Nacional / MCTI, The Ohio State University, 
Pennsylvania State University, Shanghai Astronomical Observatory, 
United Kingdom Participation Group,
Universidad Nacional Aut\'onoma de M\'exico, University of Arizona, 
University of Colorado Boulder, University of Oxford, University of Portsmouth, 
University of Utah, University of Virginia, University of Washington, University of Wisconsin, 
Vanderbilt University, and Yale University.






\bsp	
\label{lastpage}
\end{document}